\documentclass[12pt]{article}   	
\usepackage{geometry}                		
\usepackage{graphicx}				
\usepackage{amssymb}
\usepackage{amsmath}
\usepackage{epstopdf}
\usepackage{natbib}
\usepackage{hyperref}
\usepackage{setspace}
\usepackage{afterpage}
\usepackage{authblk}
\usepackage{pdflscape}
\usepackage{threeparttable}
\hypersetup{colorlinks=true, linkcolor=blue, citecolor=blue,urlcolor=blue}
\usepackage[all]{hypcap}
\usepackage{lineno}
\usepackage{authblk}
\usepackage[labelfont=bf]{caption}
\usepackage{color}

\usepackage[normalem]{ulem}  

\def\eqlbl#1{\label{eq:#1}}
\def\eqref#1{(\ref{eq:#1})}

\title{Carbon cycling and habitability of {Earth-size} stagnant lid planets}
\author[1,*]{Bradford J. Foley}
\author[1]{Andrew J. Smye}
\date{}							

\affil[1]{Department of Geosciences, Pennsylvania State University, University Park, PA, 16802}
\affil[*]{Corresponding author. E-mail: bjf5382@psu.edu. Phone: (814)-863-3591}

\begin{document}
\maketitle


Running title: Habitability of {Earth-size} stagnant lid planets



\begin{abstract}
{Models} of thermal evolution, crustal production, and CO$_2$ cycling are used to constrain the prospects for habitability of rocky planets, with Earth-like size and composition, in the stagnant lid regime. Specifically, we determine the conditions under which {such} planets can maintain rates of CO$_2$ degassing large enough to prevent global surface glaciation, but small enough {so as not to} exceed the upper limit on weathering rates {provided by the supply of fresh rock}, a situation which would lead to runaway atmospheric CO$_2$ accumulation and an inhospitably hot climate. The models show that stagnant lid planets with initial radiogenic heating rates of 100-250 TW, and with {total} CO$_2$ budgets ranging from {$\sim 10^{-2} -1$} times Earth's estimated CO$_2$ budget, can maintain volcanic outgassing rates suitable for habitability for $\approx 1-5$ {Gyrs; larger CO$_2$ budgets result in uninhabitably hot climates, while {smaller budgets} result in global glaciation.} {High radiogenic heat production rates favor habitability by sustaining volcanism and CO$_2$ outgassing longer.} Thus, the results suggest that plate tectonics may not be required for establishing a long-term carbon cycle and maintaining a stable, habitable climate. {The model is necessarily highly simplified, as the uncertainties with exoplanet thermal evolution and outgassing are large. Nevertheless, the results provide some first order guidance for} future exoplanet {missions, by predicting the age at which habitability becomes unlikely for a stagnant lid planet as a function of initial radiogenic heat budget. This prediction is powerful because both planet heat budget and age can potentially be constrained from stellar observations.}

\end{abstract}

\clearpage

\section{Introduction}
\label{sec:intro}

Thousands of exoplanets have been discovered to date, and many more are certain to follow \citep[e.g.][]{Batalha2014}, raising the question of which, if any, of these planets might be suitable hosts for life. As a result, understanding the factors that allow a planet to be habitable, in particular for surface life which can be detected by remote observations, has become an increasingly important goal. While the question of what requirements a planet must meet in order to potentially harbor life has long been a topic of research, the recent boom in exoplanet discoveries brings more urgency to this question. Theoretical ideas about planetary habitability are now potentially testable with exoplanet observations, and theoretical considerations can be used to guide astronomers searching for life to those planets that are most likely to be habitable. 

In order to host surface life, a planet must receive neither too much, nor too little, solar radiation for liquid water to be stable on the planet's surface. The focus for this study, and for most exoplanet astrobiology, is on surface life, because surface life can produce detectable signatures in a planet's atmosphere or on its surface. Life is certainly possible in oceans which lie beneath surface ice layers, such as those on Europa or Enceladus. However, such life would be difficult to detect by remote observations. The constraint that solar flux not preclude the existence of liquid surface water defines the ``habitable zone," the range of orbital distances around a star where liquid water may be stable \citep[e.g.][]{Kasting1993}. However, lying within the habitable zone does not guarantee that a planet's surface conditions will be conducive to life. {Extreme variations} in greenhouse gas concentrations can cause temperatures far in excess of what Earth-like life can tolerate {(the upper limit for known life is $\approx 400$ K; \cite{Takai2008}),} and temperatures low enough to cause global glaciation \citep[e.g.][]{walker1981,Sleep2001b,Abbot2012,Kadoya2014,Menou2015,Foley2016_review}. On Earth such variations are held in check by the long-term carbon cycle, and the stabilizing effect it has on climate \citep[e.g.][]{walker1981,berner1997}. As temperature and atmospheric CO$_2$ content increase, rates of chemical weathering increase in turn, thereby drawing CO$_2$ out of the atmosphere and acting to cool the climate. Likewise when temperature and atmospheric CO$_2$ content decrease, weathering rates slow down and volcanic CO$_2$ accumulates in the atmosphere, acting to warm the climate \citep[e.g.][]{Kump2000,Berner2004}.   

The long-term carbon cycle on Earth is facilitated in many ways by plate tectonics \citep[e.g.][]{Kasting2003}. Plate tectonics drives uplift and orogeny, which in turn spurs erosion and enhances weathering \citep[e.g.][]{West2012}. Plate tectonics also leads to the creation of continents, which increase the area of exposed land and also enhance weathering \citep[e.g.][]{Foley2015_cc}. Finally, plate tectonics also promotes long-lived CO$_2$ degassing by driving volcanism at mid-ocean ridges and island arcs, and by recycling surface CO$_2$ into the mantle such that the mantle CO$_2$ reservoir is continuously resupplied. However, whether plate tectonics is required for the long-term carbon cycle to operate, and therefore stabilize a planet's climate, is not known \citep{Foley2016_review,Lenardic2016}.  

A particularly interesting question is whether any form of climate regulation caused by chemical weathering can be maintained on a planet in the stagnant lid regime. In the stagnant lid regime the lithosphere is rigid and nearly immobile, and no subduction occurs; convection only takes place beneath the rigid lid, in the form of drip-like downwellings from the base of the lid \citep[e.g.][]{Ogawa1991,Davaille1993,slava1995}. As a result, recycling of surface material back into the mantle, in particular volatiles like carbon and water, is {seemingly} limited. The rigid, immobile surface also limits tectonic uplift, as the typical processes driving orogeny, i.e. continental collision and subduction, are no longer active. However, stagnant lid planets still experience volcanism, which can release mantle CO$_2$ to the atmosphere, create fresh rock and topography at the surface, and potentially even allow for some surface recycling through burial by lava flows \citep[e.g.][]{Pollack1987,Hauck2002,Hirschmann2008,Gillmann2014,Lenardic2016}. 

This paper sets out to test whether the long-term carbon cycle can regulate atmospheric CO$_2$ levels, and thus maintain a stable climate over geologic timescales, on a planet in the stagnant lid regime. {Previous studies have considered the ability of stagnant lid planets to maintain temperate climates, primarily by focusing on CO$_2$ degassing rates on Mars \citep[e.g.][]{Pollack1987,ONeill2007_mars,Grott2011}, a hypothetical stagnant lid Earth \citep{Tosi2017}, and exoplanets \citep{Noack2017}. However, a systematic study that considers both degassing and weathering limits to climate stability, for a range of mantle volatile and heat budgets, is lacking. We restrict our study to planets with an Earth-like size and composition. Little is known about mantle rheology, volcanism, and outgassing on planets with different bulk compositions, and hence different mantle mineralogies, than Earth, so it would be premature to model such planets. Moreover, the rheology of super-Earth mantles, even for an Earth-like composition, is also highly uncertain as super-Earth mantles reach pressures much higher than experimental studies can currently attain \citep[e.g.][]{Karato2010}. We thus focus on Earth-size planets to avoid such uncertainties. 

The potential for habitability on} stagnant lid planets {is important to assess} because they are common in the solar system; only the Earth has a mantle that convects in a plate-tectonic regime \citep[e.g.][]{Breuer2007}. Moreover, predicting whether rocky exoplanets will have plate tectonics or stagnant lids is difficult due to {our incomplete understanding} of lithospheric rheology, exoplanet heat budgets, and material properties at extremely high pressure  \citep{Valencia2007,ONeill2007,Kite2009,Korenaga2010,Stamenkovic2011,vanHeck2011,Foley2012,Lenardic2012,Noack2014}. Thus, if plate tectonics is not required for harboring life, then the search for habitable exoplanets is greatly simplified.   

\subsection{Modeling habitability of a stagnant lid planet}
\label{sec:intro_model_setup}

{In order to test whether temperate climates can be sustained on Earth-like stagnant lid planets, we couple simple models for mantle thermal evolution, CO$_2$ outgassing, weathering, and crustal growth.} The modeling focuses on two main issues: 1) if the supply of weatherable rock is sufficient for chemical weathering to balance the CO$_2$ being released by magmatism and metamorphism, and thus allow a weathering feedback to operate; and, 2) for how long, if at all, CO$_2$ outgassing rates high enough to prevent a planet's entire surface from becoming frozen over can be maintained. 

{ The first criterion is important to habitability, because the} rate at which fresh, unweathered rock is brought to the {near surface environment,} replenishing rock that has been altered via weathering, sets the ultimate upper limit on the CO$_2$ drawdown rate. In other words, the upper limit to CO$_2$ drawdown occurs when weathering immediately and completely alters all available fresh {rock}, as soon as it is supplied to the weathering zone \citep[e.g.][]{West2005,Mills2011,Foley2015_cc}, a situation known as ``supply limited" weathering \citep[e.g.][]{Stallard1983,Kump2000,Riebe2004,West2005}. On a planet where weathering becomes globally supply limited, the CO$_2$ drawdown rate only depends on the rate at which fresh rock is brought to the surface, and no longer depends on temperature or atmospheric CO$_2$ content. Thus weathering cannot adjust to balance CO$_2$ outgassing, accumulation of atmospheric CO$_2$ continues unabated, and an inhospitably hot climate develops{, as long as volcanism lasts long enough to significantly increase atmospheric CO$_2$} \citep{Foley2015_cc}. 

In contrast, when {weathering is not supply limited a feedback that stabilizes temperate surface temperatures is possible. The exact mechanism behind this feedback is unclear. In particular, the extent to which it depends on temperature as a result of weathering being limited by mineral dissolution reaction kinetics \citep[e.g.][]{Kump2000,Berner2004}, precipitation rates as a result of limitation by pore water flow rates \citep{Maher2014}, or some combination of the two, is still a topic of research. 
Given these large uncertainties involved in how silicate weathering operates, especially when extrapolated to exoplanets, we chose to simply determine when weathering is supply limited and when it is not in our modeling. 
Calculating actual climate states is beyond this study's scope. 

On Earth, physical erosion, which is greatly aided by large topographic gradients provided by mountain building and orogeny, is the primary mechanism for replenishing mineral supply for weathering on land \citep[e.g.][]{West2005,West2012}, while mid-ocean ridge volcanism supplies fresh rock on the seafloor \citep[e.g.][]{Alt1999}}. However, on a stagnant lid planet, mountain building is muted by the lack of plate collisions. We thus assume here that volcanism is the sole source of fresh, weatherable rock on a stagnant lid planet, and therefore associate volcanic eruption rates with the rate of mineral supply for weathering (as explained in more detail in \S \ref{sec:supply_lim}). 

{The second criterion is important for habitability because even with an active weathering-climate feedback, low CO$_2$ outgassing rates can induce global glaciation \citep{Kadoya2014,Haqq2016}.} We assume that CO$_2$ outgassing is a result of mantle volcanism, which releases mantle CO$_2$, and metamorphic decarbonation of the crust, originally carbonated by weathering, as it is buried by subsequent lava flows. Thus the longevity {and strength} of CO$_2$ degassing is largely controlled by the longevity {and rate} of volcanic activity. The formulation for degassing, as well as for stagnant lid thermal evolution, is given in the next section. {Note that our model results map out conditions where habitable climates are possible on stagnant lid planets, but cannot guarantee that such climates will exist in reality. Even when weathering is not supply limited, and balance between weathering and degassing can be established, inhospitably hot climates are still possible due to very high degassing rates or weakening of the feedback between weathering and climate by factors such as a lack of topography \citep[e.g.][]{Maher2014}. Moreover complexities not included in our modeling, such as additional greenhouse gases or atmospheric loss to space, could also cause uninhabitable climates to form even with sufficient CO$_2$ outgassing and fresh mineral supply rates.} 

Our model makes some important further assumptions. We assume model planets lie within the habitable zone and possess a large supply of surface water, such that weathering can occur {\citep[e.g.][]{Berner2004,Maher2014}; that is} we assume that the availability of surface water and precipitation does not limit weathering. We assume that weathering can take place both on exposed, or subaerial, land and in crust on the seafloor (i.e. seafloor weathering), as seafloor weathering has been shown recently to be an important aspect of the carbon cycle on Earth \citep{Gillis2011,Coogan2013,Coogan2015,KT2017}. As a result we do not make any explicit distinction between weathering on land or on the seafloor; all CO$_2$ released to the atmosphere is assumed to be weathered out and stored in the crust, so long as the upper bound on the weathering rate provided by the supply of weatherable rock is not exceeded. We also assume that the mantle has an oxidation state similar to Earth's mantle, such that CO$_2$ is a primary outgassing product of mantle volcanism. Such an assumption is reasonable for planets Earth-sized and larger, as mantle oxidation state is thought to largely be set by chemical reactions that occur in the lower mantle early in a planet's history, and are expected to occur on any planet large enough to have a mantle dominated by bridgmanite \citep{Wade2005}. Photolysis of water during a magma ocean stage can also contribute to mantle oxidation \citep{Hamano2013}.   

\section{Theory}
\label{sec:theory}

\subsection{Thermal evolution model}
\label{sec:thermal_evol}

We model the thermal evolution of stagnant lid planets by assuming pure internal heating, and that heat loss results from heat conducted across the lithosphere as controlled by mantle convection, and from mantle melting and magmatism. Magmatism contributes to cooling the mantle because cooling of the magma from the {temperature it is erupted or emplaced at} to the surface temperature, and the release of latent heat upon solidification, both represent a net heat loss from the mantle \citep[e.g][]{Richter1985,Bickle1986,Hauck2002,Moore2003,Nakagawa2012,Driscoll2014}. {This simple model captures the first order controls on the thermal evolution of a rocky planet and greatly simplifies the model by allowing us to focus solely on mantle evolution.} Including core heat flow would require a model for the thermal evolution of the core, adding more poorly constrained parameters to the problem. Finally, our assumptions result in a model that is conservative, in that it will give the fastest reasonable cooling history, and hence earliest shutoff to large scale volcanism and CO$_2$ degassing. Including heat flow into the mantle due to core cooling would only act to prolong volcanism (see \S \ref{sec:discussion} for further discussion). As a result, the model will produce minimum estimates for the lifetime of {potentially} habitable climates on stagnant lid planets. 
 

For magmatic heat loss, we assume that all the melt generated rises to the surface and contributes to cooling the mantle and the growth of the crust. However, only a fraction of the melt is assumed to actually erupt on the surface (an important factor for weathering, as discussed in \S \ref{sec:supply_lim}), with the rest solidifying at shallow depth; this assumption is analogous to mid-ocean ridge volcanism where only the layer of pillow lavas is erupted, with sheeted dikes and gabbros solidifying in the subsurface. As the whole melt layer is emplaced in the near surface environment, we assume all of the melt contributes to mantle heat loss. This assumption is also conservative, as it will give the maximum mantle cooling rate and thus a minimum estimate of degassing lifetime. Melt migration through a thick stagnant lid is not well understood, however, and melt could stall during its ascent through the lid, or even at the lid base. Melt stalling would slow mantle cooling and prolong volcanism, but also reduce outgassing. Test cases where some melt is assumed to stall at the base of the stagnant lid are shown in Appendix \ref{sec:melt_stall}, and give results consistent with those in the main text, where all melt rises to the surface. 

\begin{figure}
\includegraphics[width=1\textwidth]{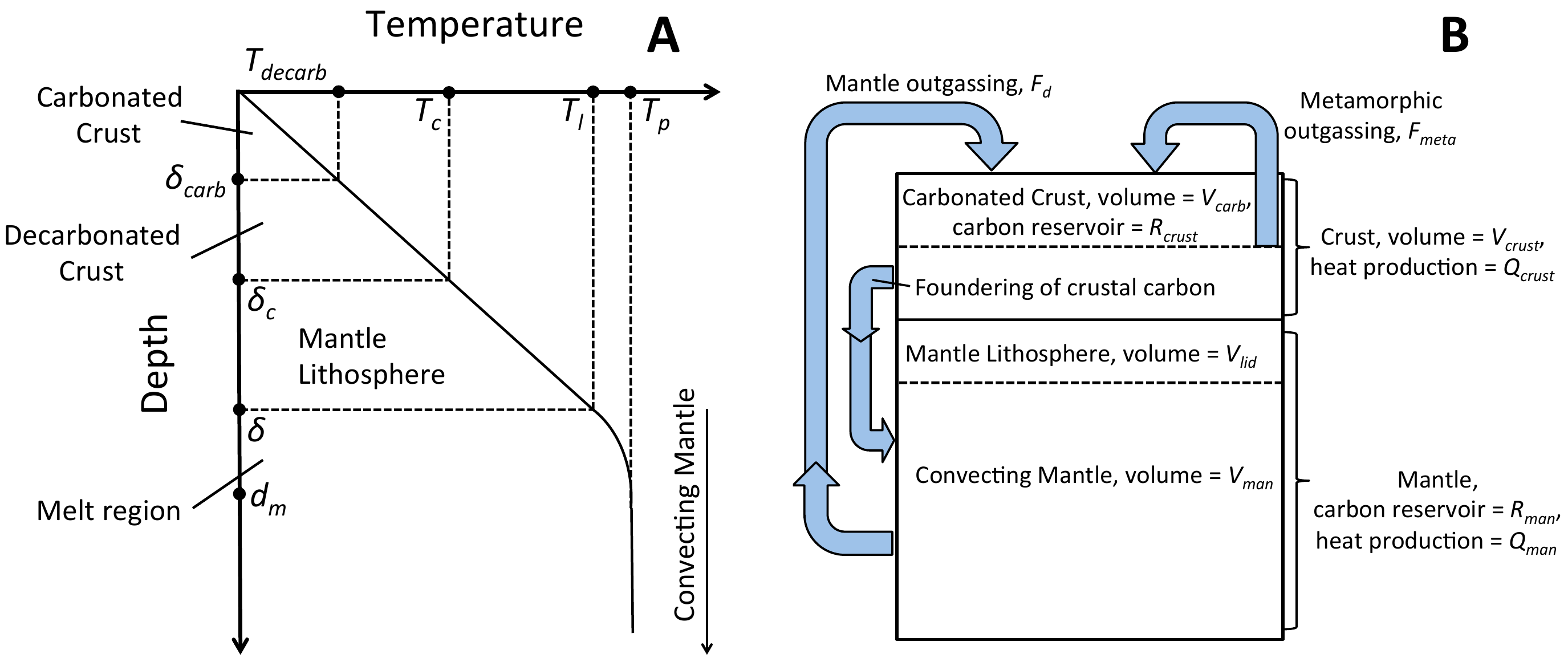}
\caption{\label {fig:schematic}  { Schematic diagram of (A) the temperature profile across the crust, lithosphere, and uppermost convecting mantle, illustrating the key layers and temperatures in the model, and (B) the major reservoirs of carbon and fluxes between these reservoirs.} }
\end{figure}

Mantle energy balance under the preceding assumptions results in the following equation for the evolution of mantle temperature \citep[e.g.][]{Stevenson1983,Hauck2002,Reese2007,Fraeman2010,Morschhauser2011,Driscoll2014}; 
\begin{equation}
\eqlbl{thermal_evol}
V_{man} \rho c_p \frac{dT_p}{dt} = Q_{man} - A_{man} F_{man} - f_m \rho_m \left (c_p \Delta T_m + L_m \right) ,
\end{equation} 
where $V_{man}$ is the volume of convecting mantle beneath the stagnant lid, $\rho$ is the bulk density of the mantle, $c_p$ is the heat capacity, $T_p$ is the potential temperature of the upper mantle, $t$ is time, $Q_{man}$ is the radiogenic heating rate in the mantle, $A_{man}$ is the surface area of the top of the convecting mantle (base of stagnant lid), $F_{man}$ is the heat flux from the convecting mantle into the base of the stagnant lid, $f_m$ is volumetric melt production, $\rho_m$ is the density of mantle melt, $L_m$ is the latent heat of fusion of the mantle, and $\Delta T_m$ is the temperature difference between the melt erupted at the surface and the surface temperature. The volume of the convecting mantle is given by $V_{man} = (4/3) \pi ((R_p-\delta)^3 - R_c^3)$, where $R_p$ is the planetary radius, $R_c$ is the core radius, and $\delta$ is the thickness of the stagnant lid. The surface area of the convecting mantle is then $A_{man} = 4 \pi (R_p-\delta)^2$. In this study we focus on Earth-sized planets, and therefore use $R_p$ equal to Earth's radius and $R_c$ equal to Earth's core radius {(see Tables \ref{tab_param} \& \ref{tab_var} for all model parameters and variables, their assumed values, and units, and Figure \ref{fig:schematic} for a sketch of the model setup.)} 

The thickness of the stagnant lid, $\delta$, evolves with time based on an energy balance model \citep[e.g.][]{Schubert1979,Spohn1991}, 
\begin{equation}
\eqlbl{delta1}
\rho c_p (T_p - T_l) \frac{d \delta}{dt} = -F_{man} - k \frac{\partial T}{\partial z} \Bigr |_{z=R_p-\delta},
\end{equation}
where $T_l$ is the temperature at the base of the lid, $k$ is the thermal conductivity, {and $z$ is height above the planet's center.} The melt heat loss terms that appear in \eqref{thermal_evol} do not appear here because we assume that heat extracted from the mantle due to volcanism is entirely lost to the surface, rather than to the stagnant lid. The mantle heat flux, $F_{man}$, and lid base temperature, $T_l$, are calculated from scaling laws {for} mantle convection \citep{Reese1998,Reese1999,Solomatov2000b,korenaga2009}: 
\begin{equation}
\eqlbl{heat_flux}
F_{man} = \frac{c_1 k(T_p - T_s)}{d} \theta^{-4/3} Ra_i^{1/3}
\end{equation}
and 
\begin{equation}
\eqlbl{T_l}
T_l = T_p - \frac{a_{rh} R T_p^2}{E_v} .
\end{equation}
Here $c_1$ is a constant, $T_s$ is the surface temperature, $d=R_p-R_c$ is the whole mantle thickness, $\theta$ is the Frank-Kamenetskii parameter, defined as $\theta = E_v (T_p - T_s)/(RT_p^2)$, which describes the strength of the temperature-dependence of mantle viscosity, $E_v$ is the activation energy for mantle viscosity, $R$ is the gas constant, $a_{rh}$ is a constant, and $Ra_i$ is the internal Rayleigh number. We use $c_1 = 0.5$ and $a_{rh} = 2.5$. Internal Rayleigh number is defined as $Ra_i = \rho g \alpha (T_p - T_s) d^3/(\kappa \mu_i)$, where $g$ is gravitational acceleration, $\alpha$ is thermal expansivity, $T_s$ is surface temperature, $\kappa$ is thermal diffusivity, and $\mu_i$ is the interior mantle viscosity, defined at temperature $T_p$. Mantle viscosity is temperature-dependent following typical flow laws for diffusion creep in olivine as $\mu_i = \mu_n \exp{(E_v/(RT_p))}$, where $\mu_n$ is a constant. We use {$E_v \approx 300$ kJ mol$^{-1}$ as the baseline value in this study \citep{karato1993}; test cases using $E_v = 200$ kJ mol$^{-1}$ and $E_v = 400$ kJ mol$^{-1}$ are shown in \S \ref{sec:sensitivity} and are consistent with our baseline model results.} We also define the reference viscosity as $\mu_r = \mu_n \exp{(E_v/(RT_r))}$, where $T_r = 1623$ K is Earth's present day mantle temperature. Using our baseline value of $\mu_n = 4 \times 10^{10}$ Pa s, $\mu_r \approx 2 \times 10^{20}$ Pa s.   

Note that in \eqref{heat_flux} the temperature difference between the mantle interior and surface, $T_p-T_s$, and whole mantle thickness, $d = R_p-R_c$, are used instead of the temperature difference between the mantle interior and base of the lid, $T_p - T_l$, and thickness of the actively convecting mantle, $d-\delta$. However, in the heat flux scaling law both the mantle thickness and temperature difference cancel out, so the equation for the heat flux to the base of the lid is independent of the definition of mantle thickness or temperature difference.  

\begin{table}
\small
\begin{threeparttable}
\caption{Model Parameters and Assumed Values}
\label{tab_param}
\begin{tabular}{c c c c}
\hline
Parameter & Meaning & Assumed value & Equation \\
\hline
$\rho$ & Mantle density & 4000 kg m$^{-3}$ & \eqref{thermal_evol} \\
$c_p$ & Heat capacity & 1250 J kg$^{-1}$ K$^{-1}$ & \eqref{thermal_evol} \\
$\rho_m$ & Melt density & 2800 kg m$^{-3}$ & \eqref{thermal_evol} \\
$L_m$ & Latent heat & 600 $\times 10^3$ J kg$^{-1}$ & \eqref{thermal_evol} \\
$R_p$ & Planet radius & 6378.1 km & below \eqref{thermal_evol} \\
$R_c$ & Core radius & 3488.1 km & below \eqref{thermal_evol} \\
$k$ & Thermal conductivity & 5 W m$^{-1}$ K$^{-1}$ & \eqref{delta1} \\
$c_1$ & Scaling law constant for $\delta$ & 0.5 & \eqref{heat_flux} \\
$d$ & Whole mantle thickness & 2890 km & \eqref{heat_flux} \\
$T_s$ & Surface temperature & 273 K & below \eqref{heat_flux} \\
$a_{rh}$ & Constant for rheological temperature scale & 2.5 & \eqref{T_l} \\
$E_v$ & Activation energy for viscosity & 300 kJ mol$^{-1}$ & \eqref{T_l} \\
$R$ & Universal gas constant & 8.314 J K$^{-1}$ mol$^{-1}$ & \eqref{T_l} \\
$g$ & Gravity & 9.8 m s$^{-2}$ & below \eqref{T_l} \\
$\kappa$ & Thermal diffusivity & $10^{-6}$ m$^2$ s$^{-1}$ & below \eqref{T_l} \\
$\alpha$ & Thermal expansivity & $3 \times 10^{-5}$ K$^{-1}$ & below \eqref{T_l} \\
$\mu_n$ & Viscosity pre-exponential factor & $4 \times 10^{10}$ Pa s & below \eqref{T_l} \\
$\mu_r$ & Reference viscosity & $2 \times 10^{20}$ Pa s & below \eqref{T_l} \\
$\gamma_{mantle}$ & Mantle adiabatic temperature gradient & $10^{-8}$ K Pa$^{-1}$ & \eqref{p_melt} \\
$\rho_l$ & Lithosphere density & 3300 kg m$^{-3}$ & \eqref{pf} \\
$(d \phi /d P)_S$ & Pressure derivative of melt fraction & $1.5 \times 10^{-10}$ Pa$^{-1}$ & \eqref{phi} \\
$\gamma_{melt}$ & Adiabatic temperature gradient of melt & $3.1 \times 10^{-8}$ K Pa$^{-1}$ & below \eqref{fm} \\
$c_2$ & Scaling law constant for $v$ & 0.05 & \eqref{vel1} \\
$c_3$ & Scaling law constant for $v$ & 0.4 & \eqref{vel2} \\
$D$ & Distribution coefficient & 0.002 & \eqref{partition} \\
$\tau_{rad}$ & Radioactive decay constant & 2.94 Gyrs & \eqref{Q_crust} \\
$A$ & Decarbonation temperature constant & $3.125 \times 10^{-3}$ K m$^{-1}$ & \eqref{T_dcarb} \\
$B$ & Decarbonation temperature constant & $835.5$ K & \eqref{T_dcarb} \\
{$D_{CO_2}$} & {Distribution coefficient for CO$_2$} & {$10^{-4}$} & {\eqref{rman} }\\
$\epsilon$ & Eruption efficiency & $0.1$ & \eqref{fws} \\
$\chi$ & Fraction of reactable elements in crust & 0.3 & \eqref{fws} \\
$\bar{m_c}$ & Molar mass of reactable elements in crust& 0.055 kg mol$^{-1}$ & \eqref{fws} \\
$T_{p0}$ & Initial mantle potential temperature & 2000 K & - \\
\end{tabular} 
\end{threeparttable}    
\end{table}

We calculate the temperature profile in the lid using a one-dimensional diffusion equation in steady-state, 
\begin{equation}
\eqlbl{lith_temp}
k \frac{\partial^2 T }{\partial z^2} = -x,
\end{equation}
where $x$ is the radiogenic heat production rate per unit volume. Within the crust, i.e. when $R_p -\delta_c \leq z \leq R_p$, $x = Q_{crust}/V_{crust} \equiv x_c$, where $\delta_c$ is the thickness of the crust, $Q_{crust}$ is the radiogenic heating rate within the crust, $V_{crust}$ is the volume of the crust, and $x_c$ is the crustal heat production rate per unit volume. Within the mantle lithosphere, i.e. when $R_p -\delta \leq z \leq R_p-\delta_c$, $x = Q_{man}/(V_{man}+V_{lid}) \equiv x_m$, where $x_m$ is the mantle heat production rate per unit volume and $V_{lid} = (4/3)\pi((R_p-\delta_c)^3-(R_p-\delta)^3)$ is the volume of the sub-crustal stagnant lid. Both the heat production, $x$, and thermal conductivity, $k$, are assumed constant {in $z$;} the difference between crustal and mantle thermal conductivity is ignored for simplicity. 

{Our temperature equation} also ignores advective heat transport, which occurs when crust newly generated by volcanism pushes the pre-existing crust down into the mantle. The Peclet Number, $Pe= \Lambda u/\kappa$ where $\Lambda$ is a length scale, $u$ is velocity, and $\kappa$ is thermal diffusivity, describes the relative importance of heat diffusion to advective transport, such that at small Peclet numbers advection is negligible. Typical eruption rates in our models are on the order of $\sim 100$ km yr$^{-3}$ (see \S \ref{sec:melt_flux}), which results in a downward velocity of crust of $\approx 6 \times 10^{-12}$ m s$^{-1}$ when crust generation is assumed to be uniform across the planet's surface. With a typical lid thickness of $\sim 100$ km (see \S \ref{sec:results}) and thermal diffusivity $\kappa = 10^{-6}$ m$^2$ s$^{-1}$, $Pe \approx 0.6$. Thus diffusion is more important than advection, but advection is not entirely negligible. Advection of cold material from the surface downward through the lid will increase the magnitude of the lid-base temperature gradient above what \eqref{lith_temp} would give. Thus, from \eqref{delta1}, the lid thickness will be underestimated by a small degree in our model {as advection is neglected. As a result, conductive cooling of the planet is overestimated by our models, which is} consistent with our other model simplifications {that act} to enhance mantle cooling. Very rapid downward advection of crust could also prevent metamorphic decarbonation (discussed further in \S \ref{sec:carbon_cycle}), as it would suppress the geotherm at depth. In \S \ref{sec:sensitivity} we explore how a lack of metamorphic decarbonation influences our results. 

Equation \eqref{lith_temp} is solved using the boundary conditions $T=T_s$ at $z=R_p$, $T=T_l$ at $z=R_p-\delta$, and by enforcing continuity of temperature and heat flux across the crust-mantle interface, i.e. at $z=R_p-\delta_c$. The lid temperature profile can be calculated analytically, meaning the temperature gradient at the base of {the} lid also has an analytical expression, and we can write 
\begin{equation}
\eqlbl{flux_lid}
{ -k \frac{\partial T}{\partial z} \Bigr |_{z=R_p-\delta} = \frac{k(T_l-T_c)}{\delta-\delta_c} - \frac{x_m(\delta-\delta_c)}{2} }
\end{equation}      
when $\delta > \delta_c$, and 
\begin{equation}
\eqlbl{flux_crust}
-k \frac{\partial T}{\partial z} \Bigr |_{z=R_p-\delta} = -\frac{x_c \delta_c}{2} + \frac{k(T_l-T_s)}{\delta_c} 
\end{equation} 
when $\delta = \delta_c$. The temperature at the base of the crust, $T_c$, is given by 
\begin{equation}
\eqlbl{T_c}
{T_c = \frac{T_s(\delta-\delta_c)+T_l\delta_c}{\delta} + \frac{x_c \delta_c^2(\delta-\delta_c) + x_m \delta_c(\delta-\delta_c)^2}{2 k \delta}.}
\end{equation} 
As described in more detail in \S \ref{sec:melting}, we assume all crust below the lid founders into the mantle, such that $\delta_c$ at all times is less than or equal to $\delta$. As the crust grows due to volcanism and $\delta_c$ approaches $\delta$, the $(\delta-\delta_c)^{-1}$ term in \eqref{flux_lid} forces the numerical solution to use very small timesteps, greatly reducing computational efficiency. To prevent this issue from severely hampering code performance, we switch to \eqref{flux_crust} when $T_c$ is within one degree Kelvin of $T_l$, which is equivalent to switching when the difference between $\delta_c$ and $\delta$ is {negligibly small}.  

\subsection{Melting and crustal evolution}
\label{sec:melting}

The melt production rate, $f_m$, is calculated in the same manner as \cite{Fraeman2010}. The pressure where melting begins, $P_i$, is 
\begin{equation}
\eqlbl{p_melt}
P_i = \frac{T_p - 1423}{120 \times 10^{-9} - \gamma_{mantle}}
\end{equation}
{using a parameterization based on the solidus for dry peridotite from \cite{Korenaga2002}. Here} $\gamma_{mantle}$ is the {average} adiabatic temperature gradient in the {upper} mantle ($\approx 10^{-8}$ K Pa$^{-1}$). Melting is assumed to stop at the base of the rigid lid, giving the final pressure of melting as
\begin{equation}
\eqlbl{pf}
P_f = \rho_l g \delta ,
\end{equation} 
{where $\rho_l$ is the average density of the crust and lithosphere (assumed to be $\rho_l = 3300$ kg m$^{-3}$).} The melt fraction, $\phi$, is 
\begin{equation}
\eqlbl{phi}
\phi = \frac{P_i - P_f}{2} \left(\frac{d\phi}{dP} \right)_S,
\end{equation}
where $(d\phi/dP)_S \approx 1.5 \times 10^{-10}$ Pa$^{-1}$. 

The melt production rate is given by the product of the melt fraction and the flux of mantle material into the melting zone. Assuming a cylindrical region of upwelling mantle \citep[as in, e.g.][]{Reese1998}, surrounded by downwelling, the volumetric flow rate of mantle into the melting region, which is equal to the flow rate of mantle out the side of the cylinder, is given by $v(d_m-\delta) \pi L$, where $v$ is the convective velocity, $d_m = P_i/(\rho_l g)$ is the depth where melting begins, and $L$ is the diameter of the cylindrical upwelling region. Dividing by the area of the circle with a radius {equal to the horizontal length of convection cells, here assumed to be $d$,} and multiplying by the surface area of the planet, the total melt production rate is 
\begin{equation}
\eqlbl{fm}
f_m = 17.8 \pi R_p v (d_m - \delta) \phi,
\end{equation}
where we have assumed that $d \approx 0.45 R_p$ as on Earth. {The formulation for melt production rate assumes that downwellings are negligibly small, such that passive upwelling makes up approximately the entire area of a convection cell, as downwellings in internally heated stagnant lid convection are narrow \citep[e.g.][]{Solomatov2000b}. One could also assume that the radius of the cylinder of upwelling mantle is only $L/2 \approx d/2$, or half the radius of convection cells. Doing so would introduce a factor if 1/2 in \eqref{fm}, which has a negligible influence on our results (see \S \ref{sec:sensitivity}). Note also that melt production, and hence crustal production and CO$_2$ outgassing, is continuous and global as long as volcanism is active, as melt is constantly generated by upwelling mantle into a global melt zone. From our thermal evolution models the depth of melting is always $< 200$ km, which is above the depth range of $\approx 250-500$ km where melt can become negatively buoyant \citep{Reese2007}. We therefore assume all melt is positively buoyant in our model.} 

The temperature difference between melt erupted at the surface and the surface temperature, $\Delta T_m = T_p - P_i \gamma - T_s$. Melt created where upwelling mantle crosses the solidus has a temperature $T_p + P_i \gamma_{mantle}$. This melt cools by adiabatic decompression as it rises to the surface, by an amount $P_i \gamma_{melt}$, where $\gamma_{melt}$ is larger than $\gamma_{mantle}$. Subtracting the adiabatic cooling from the initial temperature gives the erupted melt temperature as $T_p - P_i \gamma$, where $\gamma = \gamma_{melt} - \gamma_{mantle} = 6.766 \times 10^{-4}$ K km$^{-1}$ {using an average value of $\gamma_{melt} \approx 1$ K km$^{-1}$ \citep{Driscoll2014}.} The erupted melt temperature subtracted by the surface temperature, $T_s$, then gives $\Delta T_m$.   

The velocity, $v$, is calculated from scaling laws for internally heated stagnant lid convection \citep{Reese1998,Reese1999,Solomatov2000b,korenaga2009}, which give either 
\begin{equation}
\eqlbl{vel1}
v = c_2 \frac{\kappa}{d} \left(\frac{Ra_i}{\theta} \right)^{2/3}
\end{equation}
or
\begin{equation}
\eqlbl{vel2}
v = c_3 \frac{\kappa}{d} \left(\frac{Ra_i}{\theta} \right)^{1/2} ,
\end{equation}
where $c_2$ and $c_3$ are empirical constants determined from mantle convection simulations. We use \eqref{vel1} as the scaling law for velocity in this study; test cases using \eqref{vel2} show no significant differences in terms of mantle degassing lifetime and the conditions for supply limited weathering (see \S \ref{sec:sensitivity}). Note that mantle viscosity is assumed to be Newtonian in this study, resulting in the above scaling law exponents. 

Mantle melting and volcanism produces a crust at the surface that grows over time. If the crustal thickness, $\delta_c$, approaches that of $\delta$ as calculated from \eqref{delta1}, the crust can begin to influence convective heat flow and velocity in a way that is not captured by the scaling laws. However, crust also undergoes phase transitions as it is buried that influence its stability. In particular, basalt transforms into eclogite, which is dense relative to the underlying mantle, at a depth of $\approx 50$ km {on Earth} \citep[e.g.][]{Hacker1996}. The lower crust may therefore founder and sink into the mantle, preventing the crust from becoming so thick that it influences the thickness of the stagnant lid, and thus heat flow and velocity. The depth at which crustal foundering would occur is unknown. The basalt to eclogite transition will cause crust to become compositionally unstable at $\approx 50$ km depth, however, with a thick stagnant lid the viscosity at this depth will be large, potentially preventing foundering. {Although some studies have looked at stagnant lid convection with a basaltic crust and the eclogite transition \citep[e.g.][]{Johnson2014}, heat flow scaling laws taking crustal foundering into account have not yet been developed.} Lacking these scaling laws at the present time, we simply assume that crust delaminates when it reaches the base of the thermal lithosphere as dictated by convective instability, as in this region the viscosity contrast between lithosphere and mantle is low. As a result, we assume that lithospheric thickness follows \eqref{delta1}, and thus all crust buried to depths below $\delta$ is recycled into the mantle. This assumption has been used in other thermal evolution studies \citep[e.g.][]{Morschhauser2011}.    

With the assumption that all of the melt produced contributes to crustal growth, the evolution of crustal volume, $V_{crust}$, is
\begin{equation}
\eqlbl{crust2}
\frac{d V_{crust}}{dt} = f_m - \left(f_m-4\pi(R_p-\delta)^2\textrm{min}\left(0,\frac{d\delta}{dt}\right) \right)(\tanh{((\delta_c-\delta)20)}+1).
\end{equation} 
The hyperbolic tangent function is a mathematically convenient way to formulate a crustal loss rate that goes to zero when $\delta_c < \delta$ and equals the crustal growth rate plus the lid thinning rate when $\delta_c = \delta$. The term $4\pi(R_p-\delta)^2\textrm{min}(0,d\delta/dt)$ gives the rate at which the volume of stagnant lid is lost when $\delta$ is shrinking, and is 0 when $\delta$ is growing. The crustal thickness, $\delta_c$, is then calculated from the crustal volume as 
\begin{equation}
\eqlbl{delta_c}
\delta_c = R_p -\left( R_p^3 - \frac{3 V_{crust}}{4 \pi} \right)^{1/3},
\end{equation}
{assuming that crustal thickness is constant spatially (i.e. that melt eruption and crustal production is evenly distributed across the planet).}

Incompatible elements, including heat producing elements, are preferentially incorporated into the melt phase during mantle melting. As a result, the crust becomes enriched in heat producing elements and the mantle becomes depleted. We track the evolution of heat producing elements in the mantle and crust by assuming accumulated fractional melting \citep[e.g.][]{Fraeman2010,Morschhauser2011}, 
\begin{equation}
\eqlbl{partition}
x_{melt} = \frac{x_0}{\phi} [1 - (1-\phi)^{1/D} ] 
\end{equation}
where $x_{melt}$ is the concentration of heat producing elements in the melt, $x_0$ is the initial concentration in the solid, and $D$ is the distribution coefficient {(we use $D=0.002$ \citep{Hart1993,Hauri1994})}. The evolution of the crustal heat production rate, $Q_{crust}$, is therefore 
\begin{equation}
\begin{split}
\eqlbl{Q_crust}
\frac{d Q_{crust}}{dt} & = \frac{x_m f_m}{\phi} [1 - (1-\phi)^{1/D} ] - \\ & x_c \left(f_m-4\pi(R_p-\delta)^2\textrm{min}\left(0,\frac{d\delta}{dt}\right) \right)(\tanh{((\delta_c-\delta)20)}+1) - \frac{Q_{crust}}{\tau_{rad}}
\end{split}
\end{equation} 
and mantle heat production evolves as 
\begin{equation}
\begin{split}
\eqlbl{Q_man}
\frac{d Q_{man}}{dt} & = x_c \left(f_m-4\pi(R_p-\delta)^2\textrm{min}\left(0,\frac{d\delta}{dt}\right) \right)(\tanh{((\delta_c-\delta)20)}+1) - \\ & \frac{x_m f_m}{\phi} [1 - (1-\phi)^{1/D} ] - \frac{Q_{man}}{\tau_{rad}}.
\end{split}
\end{equation} 
The rate at which heat production is added to the crust due to melting is given by the product of the concentration of heat producing elements in the liquid and the volumetric melt production rate, $f_m$. Associating the concentration of heat producing elements in the mantle at a given time, $x_m$, with $x_0$ from \eqref{partition}, the rate of increase in $Q_{crust}$ due to melting is given by {$(x_mf_m/\phi) [1 - (1-\phi)^{1/D} ] $,} which is the first term on the right hand side of \eqref{Q_crust}. The second term on the right hand side of \eqref{Q_crust} represents recycling of crustal heat producing elements into the mantle due to crustal foundering, and the last term is the radiogenic decay of heat producing elements. A constant decay constant, $\tau_{rad} \approx 2.94$ Gyrs, is used based on an average of the four major radiogenic isotopes powering Earth's mantle, $^{238}$U, $^{235}$U, $^{232}$Th, and $^{40}$K \citep{Driscoll2014}. The evolution of mantle heat production follows from \eqref{Q_man}; heat producing elements are lost due to melting, gained from crustal foundering, and decay with time. All models start with {a chosen} initial heat production rate, $Q_0$, that is entirely within the mantle. {$Q_0$ is varied over a range of 5-250 TW in our models. For reference, typical estimates for the Earth span $\approx 70-100$ TW, when present day heating rates in the mantle and continental crust are combined and extrapolated back in time \citep[e.g.][]{korenaga2006}.} 

\begin{table}
\small
\begin{threeparttable}
\caption{Model Variables}
\label{tab_var}
\begin{tabular}{c c c}
\hline
Variable & Meaning {(units)} & Equation \\
\hline
$T_p$ & Mantle potential temperature {(K)} & \eqref{thermal_evol} \\
$t$ & Time {(s)} &\eqref{thermal_evol} \\
$V_{man}$ & Volume of convecting mantle {(m$^3$)} & \eqref{thermal_evol} \\
$A_{man}$ & Surface are of top of convecting mantle {(m$^2$)} & \eqref{thermal_evol} \\
$Q_{man}$ & Radiogenic heat production in the mantle {(W)} & \eqref{thermal_evol} and \eqref{Q_man} \\
$F_{man}$ & Mantle convective heat flux {(W m$^{-2}$)} & \eqref{thermal_evol} \\
$f_m$ & Volumetric melt production rate {(m$^3$ s$^{-1}$)} & \eqref{thermal_evol} \\
$\Delta T_m$ & Temperature difference between erupted melt and $T_s$ {(K)} &  \eqref{thermal_evol} and below \eqref{fm} \\
$\delta$ & Thickness of stagnant lid {(m)} & \eqref{delta1} \\
$T_l$ & Temperature at base of stagnant lid {(K)} & \eqref{delta1} and \eqref{T_l} \\
$\theta$ & Frank-Kamenetskii parameter {(unit-less)} & \eqref{heat_flux} \\
$Ra_i$ & Internal Rayleigh number {(unit-less)}  & \eqref{heat_flux} \\
$\mu_i$ & Interior mantle viscosity {(Pa s)}  & below \eqref{heat_flux} \\
{$x$} & {Concentration of heat producing elements {(W m$^{-3}$)}} & {\eqref{lith_temp}} \\
$V_{crust}$ & Volume of crust {(m$^3$)} & below \eqref{lith_temp} and \eqref{crust2} \\
$\delta_c$ & Thickness of crust {(m)} & below \eqref{lith_temp} and \eqref{delta_c} \\
$Q_{crust}$ & Radiogenic heat production in the crust {(W)} & below \eqref{lith_temp} and \eqref{Q_crust} \\
$V_{lid}$ & Volume of subcrustal stagnant lid {(m$^3$)} & below \eqref{lith_temp} \\
$x_m$ & Concentration of heat producing elements in the crust {(W m$^{-3}$)} & \eqref{flux_lid} \\
$x_c$ & Concentration of heat producing elements in the crust {(W m$^{-3}$)} & \eqref{flux_crust} \\
$T_c$ & Temperature at the base of the crust {(K)} & \eqref{T_c} \\
$P_i$ & Pressure where melting begins {(Pa)} & \eqref{p_melt} \\
$P_f$ & Pressure where melting ceases {(Pa)} & \eqref{pf} \\
$\delta$ & Thickness of the lithosphere {(m)} & \eqref{pf} \\
$\phi$ & Melt fraction {(unit-less)} & \eqref{phi} \\
$d_m$ & Depth where melting begins {(m)} & \eqref{fm} \\
$v$ & Convective velocity {(m s$^{-1}$)} & \eqref{vel1} and \eqref{vel2} \\
$Q_{0}$ & Initial radiogenic heat production {(W)} & below \eqref{Q_man} \\
$T_{decarb}$ & Temperature of metamorphic decarbonation {(K)} & \eqref{T_dcarb} \\
$\delta_{carb}$ & Decarbonation depth {(m)} & \eqref{d_carb} \\
$R_{man}$ & Mantle carbon reservoir {(mol)} & \eqref{rman} \\
$F_d$ & Mantle degassing flux {(mol s$^{-1}$)} & below \eqref{rman} \\
$R_{crust}$ & Crustal carbon reservoir {(mol)} & \eqref{rcrust} \\
$F_{meta}$ & Metamorphic degassing flux {(mol s$^{-1}$)} & \eqref{f_meta} \\
$C_{tot}$ & Total carbon budget {(mol)} & below \eqref{rman2} \\
\end{tabular} 
\end{threeparttable}    
\end{table}

\subsection{Carbon cycle} 
\label{sec:carbon_cycle}

In order to track how the CO$_2$ degassing rate decays with time, and whether CO$_2$ degassing overwhelms CO$_2$ drawdown via weathering, a simple model of carbon cycling between the surface and mantle is employed. Carbon is lost from the mantle during volcanism and deposited in the crust via weathering. As carbonated crust is buried by subsequent lava flows, temperature and pressure conditions are typically high enough for crustal decarbonation to occur \citep[e.g.][]{Kerrick2001b}. This metamorphic devolatilization results in an additional flux of CO$_2$ to the atmosphere, which also weathers out and is redeposited in the newly forming crust at the surface.   

Formulating the carbon cycle model as described in the previous paragraph assumes that the rate of atmospheric CO$_2$ drawdown via weathering is always equal to the rate of CO$_2$ input via volcanic and metamorphic degassing, as long as the supply limit to weathering is not exceeded. This assumption is justifiable because the dependence of the weathering rate on surface temperature and atmospheric CO$_2$ content allows a balance between weathering and degassing to be quickly established \citep{Sundquist1991,berner1997}. As the weathering rate increases with increasing temperature and partial pressure of atmospheric CO$_2$, a situation where the degassing flux is larger than the weathering flux will cause the weathering flux to increase until balance is re-established. An analogous process occurs for the situation when the weathering flux is greater than the degassing flux; cooling of the climate will cause the weathering rate to decrease until degassing and weathering come back into balance. Assuming that degassing and weathering are always in balance allows us to use a simple model that only tracks carbon in the mantle and the crust; no model for climate and atmospheric carbon evolution is required. By neglecting the atmosphere and ocean carbon reservoirs, the model used in this study will slightly overestimate the mantle and crustal carbon reservoirs. However, the atmosphere and ocean carbon reservoirs are small, so this overestimation is not significant \citep[e.g.][]{Sleep2001b}. 

\begin{figure}
\includegraphics[width=\textwidth]{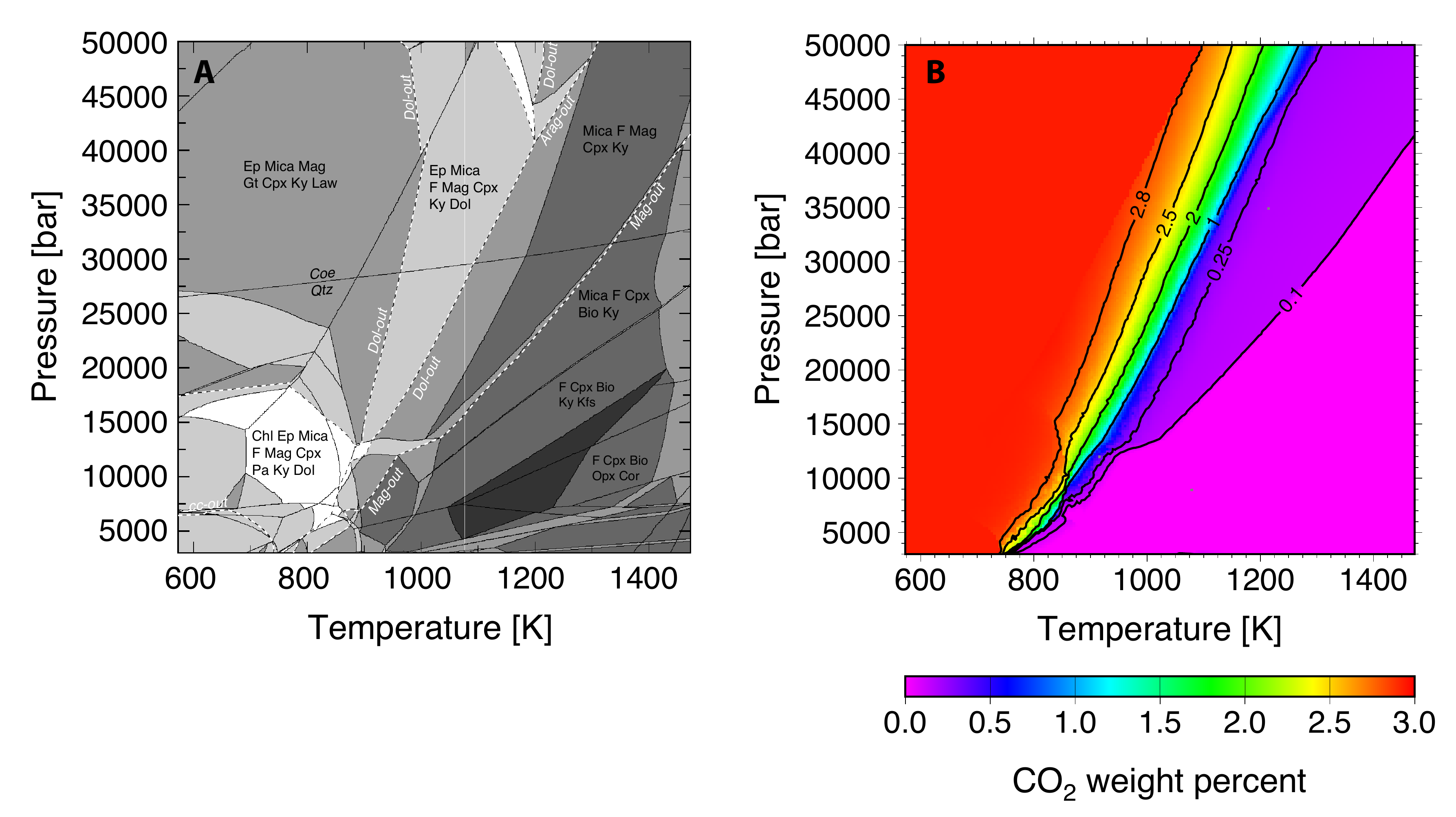}
\caption{\label {fig:phase}  Pseudosection computed for carbonate-bearing oceanic metabasalt. Panel A shows phase relations with assemblage fields shaded according to variance. Rutile, SiO$_2$, and hematite are present within each of the labeled fields. Assemblages in small fields are omitted for clarity. Dashed white lines denote discontinuous reactions involving carbonate phases. Solid black lines denote phase assemblage boundaries. Phase abbreviations are as follows: Ep = epidote, Mag = magnesite, Gt = garnet, Cpx = clinopyroxene, Ky = kyanite, Law = lawsonite, Dol = dolomite, F = H$_2$O-CO$_2$ fluid, Bio = biotite, Mica = white mica, Kfs = K-feldspar, Cor = corundum, Pa = paragonite, Cc = calcite, Qtz = quartz, Coe = coesite. Panel B shows false color contour map of weight \% CO$_2$ in the phase assemblage section presented in panel A.  }
\end{figure}

The depth at which crustal decarbonation occurs determines whether carbonated crust is recycled into the mantle by convective foundering, or whether the crust degasses its carbon back to the atmosphere before it founders. The crustal decarbonation depth, $\delta_{carb}$, is determined based on the topology of a P-T phase diagram section calculated for a carbonate-bearing oceanic metabasalt \citep[e.g.][]{Staudigel1989,Kerrick2001b} (see Figure \ref{fig:phase}). These calculations were performed using Perple{\_}X 6.6.6 \citep{Connolly2005}, in combination with the Holland and Powell DS622 thermodynamic database \citep{Holland2011}. Details of the mineral activity-composition models used are provided in Appendix \ref{sec:appendix}. Decarbonation reactions have positive Clapeyron slopes across the depth-interval relevant to our model; slopes become steeper at pressures above $\approx 30$ kbar. The bulk of metamorphic decarbonation ($>2.8$ to $<0.25$ wt. \% CO$_2$) occurs across a narrow ($\approx 100 - 200$ K) temperature interval, chiefly by the breakdown of dolomite (CaMg(CO$_3$)$_2$). Magnesite is the principal host for CO$_2$ at temperatures less than 1000 K; calcite and aragonite have small stability fields. The effect of these metamorphic phase relations is incorporated into the model by approximating the pressure and temperature dependence of the 0.25 wt. \% CO$_2$ contour with a simple linear fit. 

The temperature at which decarbonation occurs is then:
\begin{equation}
\eqlbl{T_dcarb}
T_{decarb} = A (R_p-z) + B
\end{equation} 
where $A=3.125 \times 10^{-3}$ K m$^{-1}$, $B=835.5$ K, and $T_{decarb}$ is temperature in kelvins. Pressure is converted into depth by assuming a lithospheric density of $\rho_l = 3300$ kg m$^{-3}$. From \eqref{lith_temp}, the temperature profile in the crust is 
\begin{equation}
\eqlbl{T_lid}
T = T_s - \frac{x_c }{2k} (R_p -z)^2 +  \frac{x_c \delta_c}{2k} (R_p -z) + \frac{(T_c-T_s)}{\delta_c}(R_p-z). 
\end{equation}
The height where decarbonation occurs is found by equating \eqref{T_dcarb} and \eqref{T_lid}. The decarbonation depth, $\delta_{carb}$, occurs at $z = R_p - \delta_{carb}$, and is therefore given by  
\begin{equation}
\begin{split}
\eqlbl{d_carb}
\delta_{carb} = & \frac{\delta_c}{2} + \frac{k(T_c-T_s)}{\delta_c x_c} - \frac{Ak}{x_c} - \\ & \frac{k}{x_c} \sqrt{\left(\frac{x_c \delta_c}{2k} + \frac{T_c-T_s}{\delta_c} - A \right)^2 + \frac{2 x_c}{k}(T_s - B)}.
\end{split}
\end{equation} 

When $\delta_{carb} < \delta$, crust decarbonates before being recycled into the mantle, and the evolution of the mantle carbon reservoir, $R_{man}$, is given by
\begin{equation}
\eqlbl{rman}
{\frac{dR_{man}}{dt} = -\frac{f_m R_{man} [1 - (1-\phi)^{1/D_{CO_2}} ]}{\phi (V_{man}+V_{lid})},}
\end{equation} 
where $R_{man}$ has units of moles \citep[e.g.][]{TajikaMatsui1992,Franck1999,Sleep2001b,Driscoll2013,Foley2015_cc}. {As crust decarbonates before being recycled into the mantle, the mantle only loses carbon over time when $\delta_{carb} < \delta$.} The right hand side of \eqref{rman} is defined as the mantle degassing flux, 
\begin{equation}
{ F_d = \frac{f_m R_{man} [1 - (1-\phi)^{1/D_{CO_2}} ]}{\phi (V_{man}+V_{lid})}, }
\end{equation}
and gives the rate at which CO$_2$ is outgassed from mantle to atmosphere, where weathering will deposit it in the crust. CO$_2$ is incompatible and is therefore preferentially lost from the solid to the melt, analogous to the heat producing elements. We use a distribution coefficient for CO$_2$ of $D_{CO_2} = 10^{-4}$ \cite[e.g.][]{Hauri2006}. As we assume that all the melt produced travels to the surface or near surface, we also assume that all of the melt produced contributes to degassing of CO$_2$ to the atmosphere. Assuming that only a fraction of the melt degasses its CO$_2$ to the atmosphere, with the rest being trapped in the crust, does not significantly change our overall results (see Appendix \ref{sec:melt_stall}). 

The evolution of the crustal carbon reservoir, $R_{crust}$, then follows as
\begin{equation}
\eqlbl{rcrust}
{ \frac{dR_{crust}}{dt} = \frac{f_m R_{man} [1 - (1-\phi)^{1/D_{CO_2}} ]}{\phi (V_{man}+V_{lid})}.}
\end{equation} 
Metamorphic decarbonation results in no net change in the crustal carbon reservoir because carbon degassed from the base of the crust is immediately (on geologic timescales) redeposited at the top of the crust via weathering. Volcanic degassing of the mantle therefore causes a net transfer of carbon from the mantle to the crust over the lifetime of a stagnant lid planet. Although not important for the evolution of the carbon crustal reservoir, the metamorphic degassing flux is important for climate, and is given as 
\begin{equation}
\eqlbl{f_meta}
F_{meta} = \frac{R_{crust}f_m}{2 V_{carb}}(\tanh{((\delta_c-\delta_{carb})20)}+1) 
\end{equation}
where $V_{carb}$ is the volume of carbonated crust, $V_{carb} = (4/3) \pi (R_p^3 - (R_p - \delta_{carb})^3)$, and, as in \eqref{crust2}, a hyperbolic tangent function is used as a mathematically convenient way to parameterize a step-function like change in crustal CO$_2$ around the decarbonation depth. Here a factor of 1/2 is included, so that the metamorphic degassing flux goes to $R_{crust} f_m/(V_{carb})$ when $\delta_c > \delta_{carb}$ and to 0 when $\delta_c < \delta_{carb}$. {Metamorphic CO$_2$ outgassing is assumed to be complete and to occur uniformly across the planet's surface, as we assume that crustal production is spatially uniform. The effect of incomplete metamorphic outgassing is explored in Appendix \ref{sec:melt_stall}, and a case where metamorphic decarbonation is ignored entirely is shown in \S \ref{sec:sensitivity}. }  

When $\delta_c < \delta_{carb}$, the following equations for $R_{crust}$ and $R_{man}$ are used: 
{
\begin{equation}
\begin{split}
\eqlbl{rcrust2}
\frac{dR_{crust}}{dt} & =\frac{f_m R_{man} [1 - (1-\phi)^{1/D_{CO_2}} ]}{\phi (V_{man}+V_{lid})}- \\ & \frac{R_{crust}}{V_{crust}}\left(f_m-4\pi(R_p-\delta)^2\textrm{min}\left(0,\frac{d\delta}{dt}\right) \right)(\tanh{((\delta_c-\delta)20)}+1)
\end{split}
\end{equation} 
\begin{equation}
\begin{split}
\eqlbl{rman2}
\frac{dR_{man}}{dt} & = -\frac{f_m R_{man} [1 - (1-\phi)^{1/D_{CO_2}} ]}{\phi (V_{man}+V_{lid})}+ \\ & \frac{R_{crust}}{V_{crust}}\left(f_m-4\pi(R_p-\delta)^2\textrm{min}\left(0,\frac{d\delta}{dt}\right) \right)(\tanh{((\delta_c-\delta)20)}+1).
\end{split}
\end{equation}}
In this case, crust buried beneath the stagnant lid base will founder into the mantle, returning surface carbon to the mantle. In practice $\delta_{carb}$ is usually found to be less than $\delta_c$ while volcanism is active, so models typically follow equations \eqref{rman}-\eqref{rcrust} rather than equations \eqref{rcrust2}-\eqref{rman2}. The sum of the mantle and crustal carbon reservoirs is the total carbon budget, $C_{tot} = R_{man} + R_{crust}$, which is conserved. All models start with the entire carbon budget residing in the mantle and degassing to the surface over time.  

\section{The ``hot" limit: supply limited weathering}
\label{sec:supply_lim}   

As discussed in \S \ref{sec:intro_model_setup}, the feedback between silicate weathering and climate that promotes habitable surface conditions only operates {if the supply of fresh rock does not limit the weathering rate} \citep[e.g.][]{West2012,Foley2016_review}. 
To formulate the global supply limit to weathering on a stagnant lid planet, we assume that volcanism is the dominant source of weatherable rock at the surface, and therefore write 
\begin{equation}
\eqlbl{fws}
F_{w_s} = \frac{\epsilon f_m \chi \rho_l}{\bar{m}_c},
\end{equation}  
where $F_{w_s}$ is the supply limited weathering flux (in units of mol yr$^{-1}$), $\epsilon$ is the eruption efficiency, the fraction of melt produced that erupts at the surface, $\chi$ is the fraction of reactable elements in the crust, and $\bar{m}_{c}$ is the average molar mass of the crust. Assuming a basaltic crust, $\chi = 0.3$ and $\bar{m}_c = 0.055$ kg mol$^{-1}$. {The above formulation for the supply limited weathering flux assumes complete carbonation of erupted basalt, which is the ultimate upper bound on silicate weathering. In practice carbonation can limit the ability of pore water to reach fresh basalt and continue CO$_2$-rock reaction; pore space can be clogged by formation of carbonated minerals and reaction products can coat fresh basalt surfaces \citep[e.g.][]{Kump2000,Kelemen2011,Beinlich2012}. As a result the true supply limit to weathering may be less than our estimate based on complete carbonation. In section \S \ref{sec:sensitivity} we discuss the impact a lower supply limit to weathering has on our results. } 

\begin{figure}
\includegraphics[width=1\textwidth]{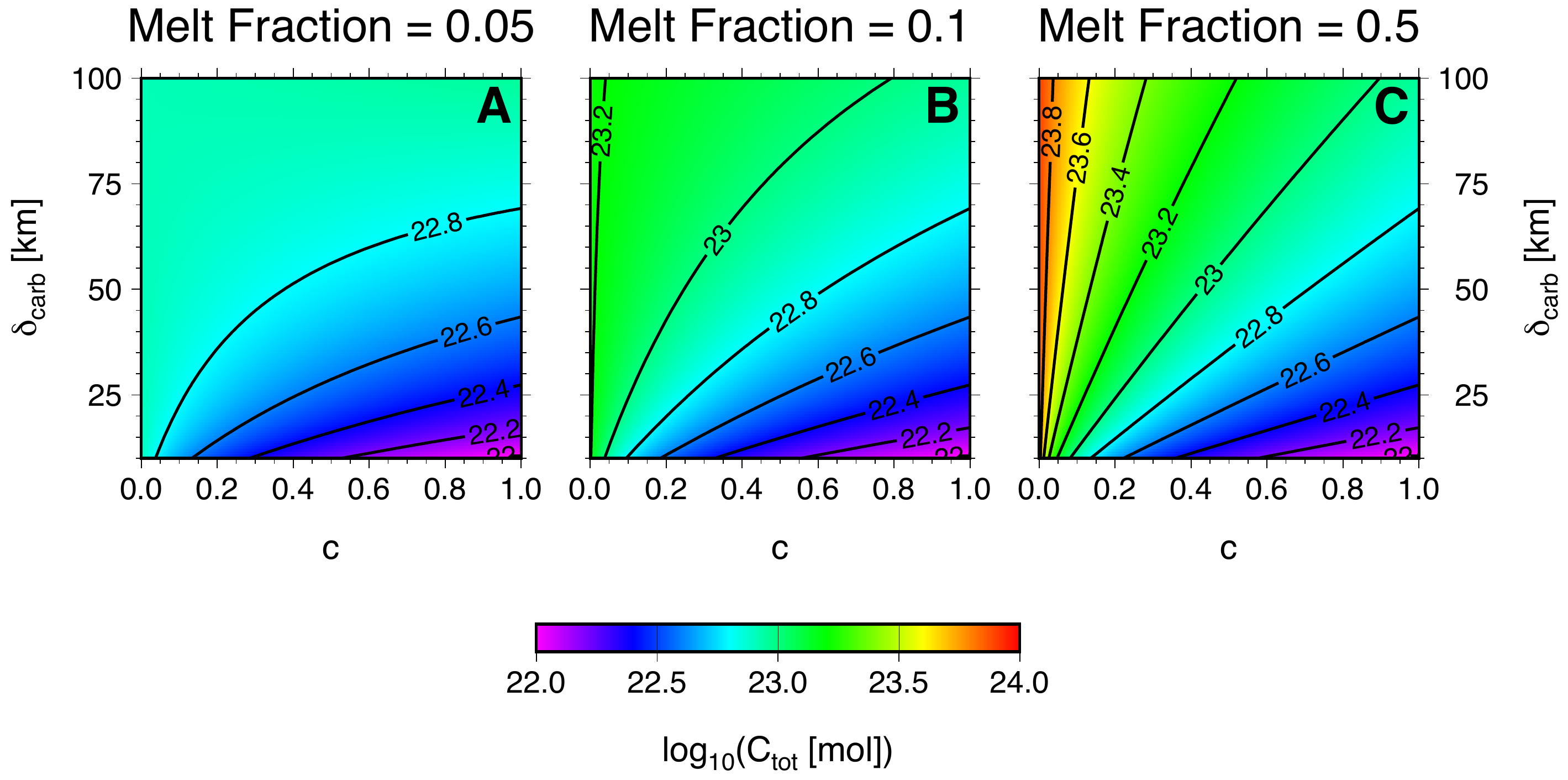}
\caption{\label {fig:C_tot_crit}  { The critical CO$_2$ budget where weathering will be supply limited ($C_{tot_{crit}}$) as a function of the fraction of the total CO$_2$ budget residing in the crust ($c$), and the decarbonation depth for the crust ($\delta_{carb}$), for $\phi = 0.05$ (A), $\phi=0.1$ (B), and $\phi=0.5$ (C).} }
\end{figure} 

Supply limited weathering prevails when $F_d + F_{meta} \geq F_{w_s}$. Thus the threshold where supply limited weathering begins is given by $F_d + F_{meta} = F_{w_s}$, and the critical mantle and crustal CO$_2$ contents where weathering is supply limited are given by 
\begin{equation}
\eqlbl{supply_lim}
{ \frac{R_{man_{crit}}}{\phi (V_{man}+V_{lid})} +  \frac{R_{crust_{crit}}}{V_{carb}} = \frac{\epsilon \chi \rho_l}{\bar{m}_c} .}
\end{equation} 
{ While volcanism is active melt fractions are typically on the order of 0.1. For $\phi >> D_{CO_2}$, $1-(1-\phi)^{(1/D_{CO_2})} \approx 1$, and the mantle degassing flux reduces to $f_m R_{man}/(\phi [V_{man} + V_{lid}])$; this simplified expression for the degassing flux is used in \eqref{supply_lim}.} There are two end-member scenarios: one where the planet's carbon is predominantly in the mantle and one where {it is} in the crust. If $R_{man} >> R_{crust}$, \eqref{supply_lim} reduces to an expression for the critical mantle CO$_2$ content for supply limited weathering: {$R_{man_{crit}} = \phi \epsilon \chi \rho_l (V_{man}+V_{lid})/\bar{m_c}$.} Assuming an eruption efficiency of 0.1 \citep{Crisp1984,White2006} and that the volume of the crust is negligible such that $V_{man} + V_{lid} = (4/3)\pi(R_p^3 - R_c^3)$, {$R_{man_{crit}} \approx 8 \times 10^{22} - 8 \times 10^{23}$ mol for melt fractions ranging from $\phi = 0.05$ to 0.5.} As crustal volume is negligible compared to mantle volume for a crust on the order of 100 km thick, the typical thickness produced by our thermal evolution models, we ignore crustal volume in calculating $V_{man} + V_{lid}$ for the remainder of this section {(crustal volume is not neglected in the thermal evolution models presented in \S \ref{sec:degass})}. 

\cite{Sleep2001b} estimate $\approx 7 \times 10^{21} - 2\times 10^{22}$ mol of CO$_2$ in Earth's mantle, so a planet would need to have a mantle CO$_2$ budget {$\approx 4 - 110$} times larger than Earth's for supply limited weathering to prevail due to mantle degassing. { At melt fractions of $\phi < 0.05$ even smaller mantle CO$_2$ budgets would lead to supply limited weathering, as CO$_2$ becomes increasingly concentrated in the melt due to its incompatible nature. However, melt fractions are typically above $0.05$ in our models while volcanism is active; lower melt fractions are only seen just before large scale volcanism ceases. At this time outgassing rates are too low and short lived to lead to any significant atmospheric CO$_2$ buildup. Thus the threshold for supply limited weathering due to mantle volcanism in practice is the range of $\approx 8 \times 10^{22} - 8 \times 10^{23}$ mol given above. Metamorphic outgassing, on the other hand, can result in a lower threshold for supply limited weathering. If} carbon is concentrated in the crust (i.e. if $R_{crust} >> R_{man}$): $R_{crust_{crit}} = \epsilon \chi \rho_c V_{carb}/\bar{m_c}$. With $\delta_{carb} = 10-100$ km, $R_{crust_{crit}} \approx 10^{22} - 10^{23}$ mol. The same amount of carbon results in a higher degassing rate when it is concentrated in the crust versus in the mantle, {for melt fractions of order 0.1,} due to the much larger volume of the mantle. Thus supply limited weathering will occur at a lower total planetary CO$_2$ budget when the CO$_2$ predominately resides in the crust. 


The tradeoffs between how carbon is distributed between crust and mantle, the depth of decarbonation, { and the melt fraction, $\phi$,} can be illustrated by introducing, $c$, the fraction of a planet's total carbon budget, $C_{tot}$, that resides in the crust. Thus $R_{crust} = c C_{tot}$, $R_{man} = (1-c)C_{tot}$, and 
\begin{equation}
{C_{tot_{crit}} = \frac{\epsilon \chi \rho_c \phi (V_{man}+V_{lid})V_{carb}}{\bar{m_c}(V_{carb}(1-c)+c \phi (V_{man}+V_{lid}))} }
\end{equation} 
where $C_{tot_{crit}}$ is the critical total planetary CO$_2$ budget above which weathering will be supply limited. Figure \ref{fig:C_tot_crit} shows that either increasing the fraction of carbon in the crust or decreasing the decarbonation depth lowers $C_{tot_{crit}}$. { Meanwhile decreasing $\phi$ mutes the influence of $C_{tot_{crit}}$ on the fraction of carbon in the crust, because at lower melt fractions the threshold for supply-limited weathering due to volcanism is smaller and closer to that from metamorphic outgassing.} During planetary evolution, {$c$, $\delta_{carb}$, and $\phi$ all} change {over time}; $c$ will increase as carbon is degassed from the mantle and deposited in the crust, $\delta_{carb}$ is inversely proportional to mantle temperature and will typically grow with time, {and $\phi$ will decrease with time}. A full thermal evolution and carbon cycle model is {therefore} needed to follow how $R_{crust}$, $\delta_{carb}$, {and $\phi$} evolve, and determine whether or not supply limited weathering {prevails; this is done in \S \ref{sec:degass}. 

However, simply checking if the total degassing flux (i.e. the sum of the metamorphic and volcanic degassing fluxes) ever exceeds the supply-limited weathering flux can give misleading results about whether inhospitably hot climates will develop in practice. In particular, at low melt fractions mantle degassing can exceed the weathering supply limit as a result of the high concentration of CO$_2$ in the melt. In such situations, though, the total outgassing rate is low, on the order of $10^{10}-10^{11}$ mol yr$^{-1}$ or less in our models, and it would take billions of years for a thick CO$_2$ atmosphere to form. Mantle volcanism with a low melt fraction does not persist for such long timescales because it only occurs when volcanism is close to ceasing entirely. We thus define the threshold for supply-limited weathering to form an uninhabitable hothouse climate as $F_d + F_{meta} \geq F_{w_s} + F_{offset}$. 
We set $F_{offset}=10^{14}$ mol yr$^{-1}$, as a net CO$_2$ input to the atmosphere of $10^{14}$ mol yr$^{-1}$ will cause 100 bar of CO$_2$ ($\approx 10^{22}$ mol) to accumulate in 100 Myrs. The boundary for the onset of supply limited weathering shown in our model results below (\S \ref{sec:degass}) was found to be insensitive to $F_{offset}$ over the range $F_{offset}=10^{13}-10^{15}$ mol yr$^{-1}$. When supply limited weathering prevails, either as a result of mantle or metamorphic outgassing, it typically persists for the lifetime of volcanism. In principle as the mantle cools and $\delta_{carb}$ grows weathering could transition back {out of the supply limited regime;} however, this is not seen in our models as $\delta_{carb}$ typically stays low (order of 10 km) during the lifetime of volcanism. }


\section{The ``cold" limit: Longevity of CO$_2$ degassing}
\label{sec:degass}

{As discussed in \S \ref{sec:intro_model_setup}, the other requirement for habitability we consider is whether degassing rates are high enough to prevent global surface glaciation; this threshold is calculated using climate models and by assuming that atmospheric CO$_2$ is set by a balance between degassing and weathering \citep[e.g.][]{Kadoya2014,Menou2015,Batalha2016}.} For an Earth-like planet, \cite{Kadoya2014} estimate that a degassing flux approximately equal to Earth's present day degassing flux is needed to keep the surface temperature above freezing for solar luminosities of $\approx 0.7-0.8 S^*$ (where $S^*$ is the present day solar luminosity), while a degassing flux of $\sim 10$ \% the present day is sufficient for solar luminosities of $\approx S^*$. \cite{Haqq2016} find that a degassing flux of $\sim 10$ \% present day Earth's allows for warm climates even for Archean solar luminosities.  
To be conservative, we take the estimates from \cite{Kadoya2014} as reasonable limits on the degassing flux necessary for preventing a snowball climate, and calculate how long CO$_2$ degassing rates can be maintained at both 10 and 100 \% of Earth's present day {total} degassing flux of $\approx 6 \times 10^{12}$ mol yr$^{-1}$ \citep{Marty1998}.  

Snowball climate states triggered by low degassing rates may not be permanent. It is possible that the climate will oscillate between globally frozen and ice free states, as weathering shuts down during a snowball phase. With weathering shut down, volcanism replenishes atmospheric CO$_2$, triggering melting and a short-lived warm climate phase. During the ice-free warm climate phase, precipitation and weathering ensue, plunging the climate back into a snowball state \citep[e.g.][]{Menou2015}. However, the period of time spent under glaciated conditions is much longer than the time spent at warm conditions during these oscillations \citep{Kadoya2014,Menou2015,Batalha2016}. For simplicity we will consider that snowball climates prevail below the degassing thresholds set in the previous paragraph, even though warm climates may exist for brief durations. 

\subsection{Thermal evolution model results}
\label{sec:results}


\begin{figure}
\includegraphics[width=0.6 \textwidth]{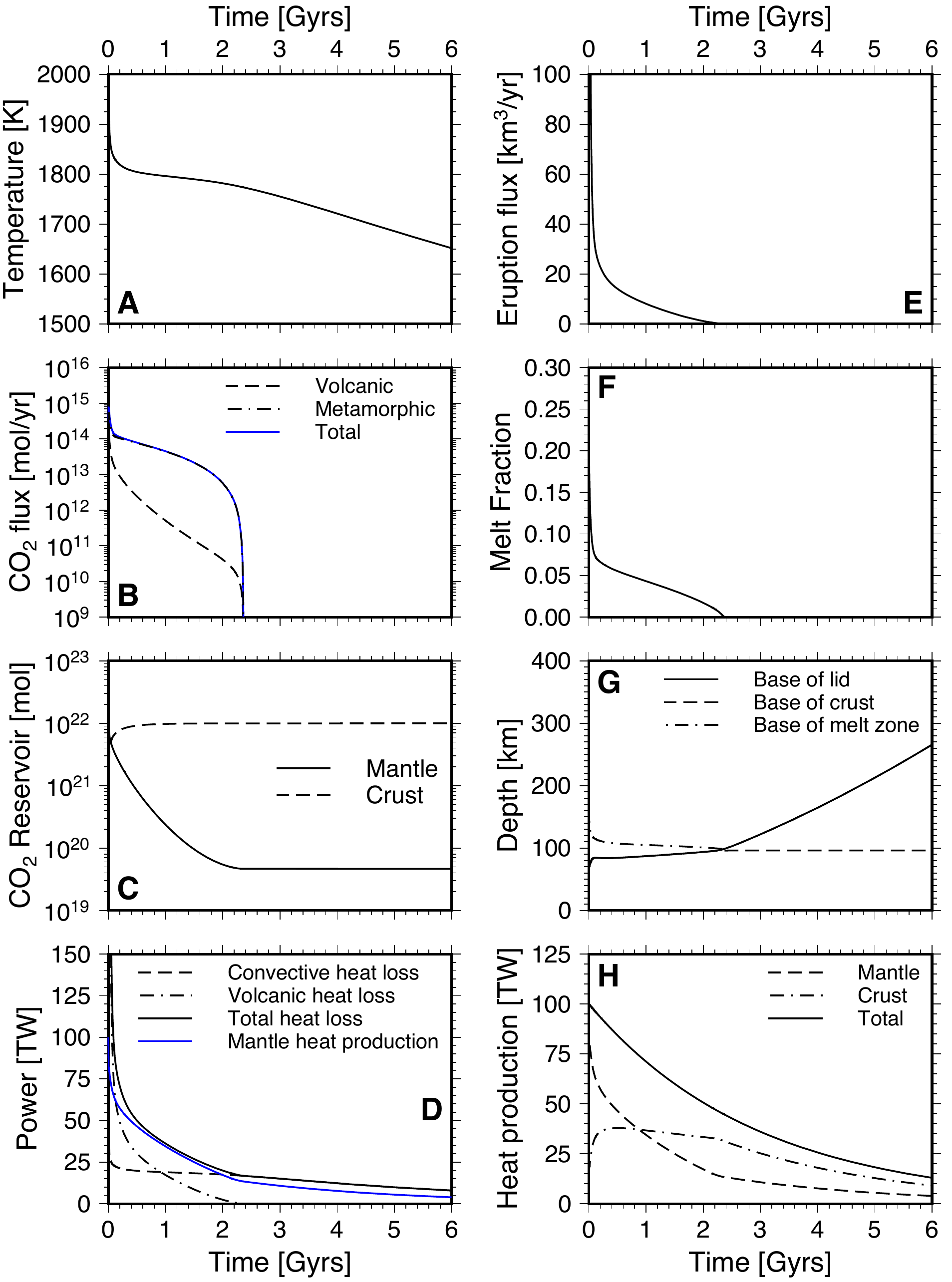}
\caption{\label {fig:evol} Evolution of: (A) mantle potential temperature; (B) mantle volcanic CO$_2$ degassing flux (dashed black line), crustal metamorphic degassing flux (dot-dashed black line), and total CO$_2$ degassing flux (solid blue line); {(C) mantle CO$_2$ reservoir (solid line) and crustal CO$_2$ reservoir (dashed line);} (D) convective mantle heat loss (dashed black line), volcanic mantle heat loss (dot-dashed black line), total mantle heat loss (solid black line) and mantle heat production (solid blue line); (E) surface volcanic eruption flux; { (F) melt fraction;} (G) thickness of the stagnant lid (solid black line), crust (dashed black line), {and base of the melt zone (dot-dashed black line)}; and (H) mantle heat production (dashed black line), crustal heat production (dot-dashed black line), and total heat production (solid black line). The thermal evolution model uses $Q_0 = 100$ TW, $T_{p_0} = 2000$ K, and $C_{tot} = 10^{22}$ mol, with the other model parameters given in Table \ref{tab_param}. }
\end{figure}  

A typical model evolution is shown in Figure \ref{fig:evol}. The model uses baseline parameter values (see Table \ref{tab_param}), an initial mantle temperature of  $T_{p_0} = 2000$ K, initial radiogenic heating rate of $Q_0 = 100$ TW, and total CO$_2$ content of $C_{tot} = 10^{22}$ mol. The initial conditions result in a low convective heat loss, less than the heat production rate, and high magmatic heat loss, far in excess of internal heat production. As a result, the mantle rapidly cools until magmatic and convective heat loss approximately balance mantle heat production. Note that without magmatic heat loss, the mantle would have initially warmed until convective heat loss became large enough to balance mantle heat production on its own. {Including magmatic heat loss therefore results in lower mantle temperatures because it increases total heat loss.} Also during this early phase of magmatic dominated cooling, a thick crust forms and becomes enriched in heat producing elements; $Q_{mantle}$ falls from 100 TW to $\approx 50$ TW within $\approx 200$ Myrs, primarily due to partitioning of heat producing elements into the crust. {Eventually the radiogenic heating rate decays to the point where convective heat loss on its own can balance heat production.} As a result, the mantle cools more rapidly and volcanism {quickly} ceases. Convective heat loss then tracks mantle radiogenic heating for the remainder of the evolution and the more rapid mantle cooling rate is sustained. The CO$_2$ degassing flux is high throughout the lifetime of volcanism due primarily to metamorphic degassing of the carbonated crust. { Metamorphic degassing dominates because CO$_2$ is more concentrated in the crust than in the mantle, and because the mantle CO$_2$ reservoir is quickly depleted by volcanism and outgassing.} CO$_2$ degassing is larger than Earth's present day degassing rate for {$\approx 2$ Gyrs, nearly the entire lifetime of volcanism.} Transferring carbon from the mantle to the crust via volcanism, and continually burying this crust through the metamorphic decarbonation depth, is thus an effective means of sustaining high CO$_2$ degassing rates. 

\begin{figure}
\includegraphics[width=\textwidth]{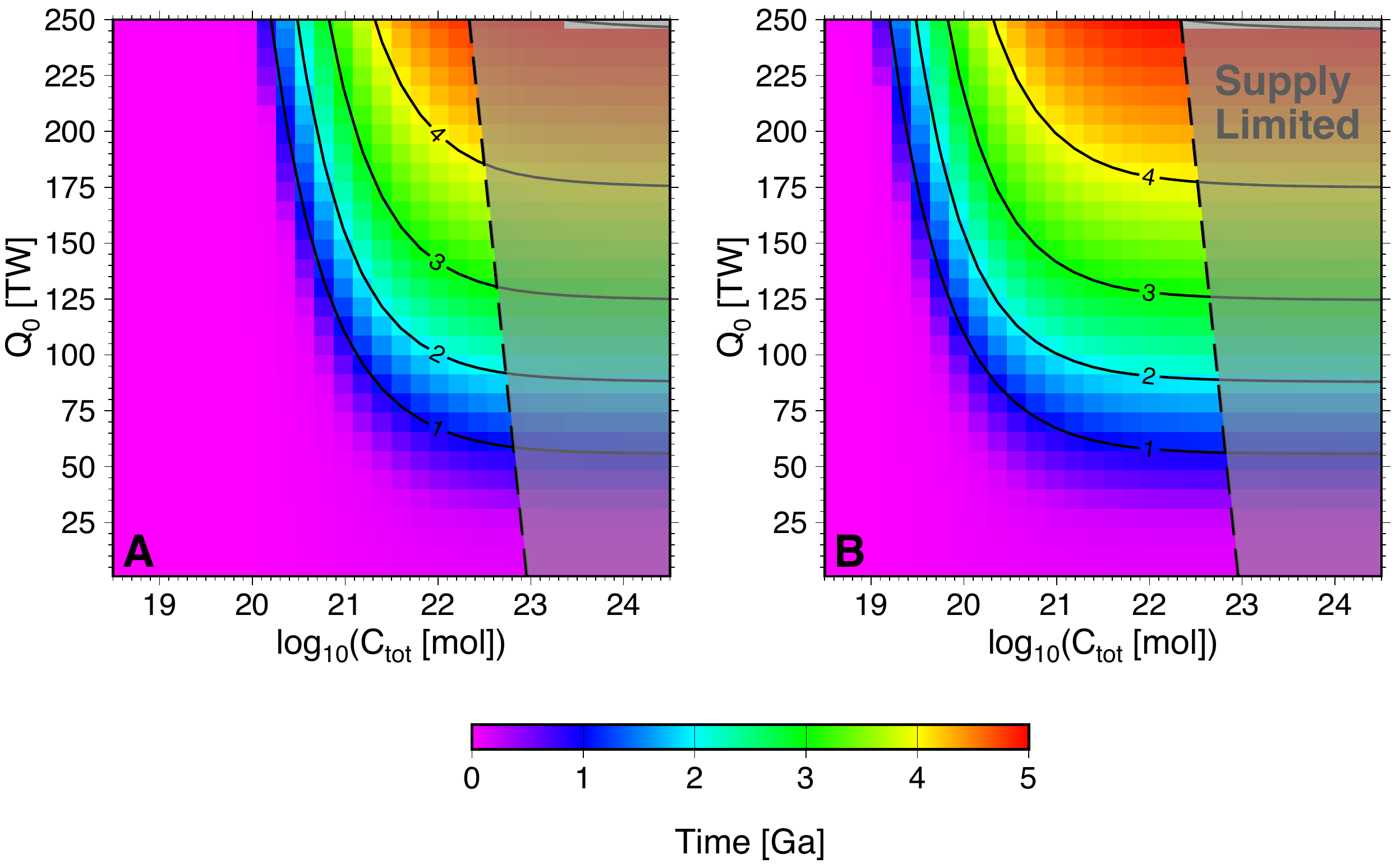}
\caption{\label {fig:fd_time1}  Time when the total degassing flux, $F_d+ F_{meta}$, falls below Earth's present day degassing flux (A), and below 10 \% of Earth's present day degassing flux (B). Labeled contours give this time in billions of years. To the right of the dashed line {(shaded region)} weathering will be supply limited assuming an eruption efficiency of 0.1 (see \S \ref{sec:supply_lim}).}
\end{figure}  

The planetary CO$_2$ budget and radiogenic heating rate are the two most important factors governing how long substantial degassing rates can be maintained on a stagnant lid planet. High radiogenic heating budgets promote long-lived volcanism and crustal burial through the decarbonation depth, while a large initial CO$_2$ content allows high degassing rates to be maintained{, primarily by metamorphic CO$_2$ degassing,} even as volcanism wanes {(i.e. with a larger CO$_2$ budget more CO$_2$ is deposited in the crust, such that degassing rates remain high even at low rates of melt production and crustal burial).} The influence of $C_{tot}$ and $Q_0$ on the longevity of CO$_2$ degassing is summarized in Figure \ref{fig:fd_time1}A, which shows the time when the total degassing flux drops below Earth's present day degassing flux, and Figure \ref{fig:fd_time1}B, which shows when the total degassing flux drops below 10 \% of Earth's present day degassing flux. As expected, both a larger radiogenic heating rate and a larger planetary CO$_2$ budget lead to longer-lived CO$_2$ degassing. The lifetime of volcanism is not a function of planetary CO$_2$ budget, and thus contours for the time when volcanism ends would form horizontal lines in Figures \ref{fig:fd_time1}A \& B. Moving to large planetary CO$_2$ budgets (i.e. towards the right edge of Figures \ref{fig:fd_time1}A \& B), the contours asymptotically approach horizontal lines because the end of volcanism, rather than the supply of CO$_2$, becomes the dominant factor determining the longevity of {adequate} degassing. Moving towards lower CO$_2$ budgets, the contours curve towards vertical lines, as the planetary CO$_2$ content becomes the primary factor controlling when degassing rates above our designated thresholds end.  Thus even with a high radiogenic heating budget, a planet with a low CO$_2$ inventory will quickly become globally frozen, even though volcanism can continue for billions of years.  

Although large CO$_2$ budgets are beneficial for promoting warm climates, there is an upper limit on the CO$_2$ inventory for habitability provided by the transition to supply limited weathering discussed in \S \ref{sec:supply_lim}. The critical planetary CO$_2$ inventory where weathering becomes supply limited was calculated from the thermal evolution models { as explained in \S \ref{sec:supply_lim}; this is found to occur when $C_{tot} \approx 10^{23}$ mol at $Q_0 = 5$ TW and $C_{tot} \approx 2 \times 10^{22}$ mol at $Q_0 = 250$ TW.} The relationship between $C_{tot_{crit}}$ and $Q_0$ is approximately linear. Larger heat production rates shift the boundary { where weathering becomes supply limited} to lower values of $C_{tot}$ (i.e. making it easier for a planet to experience supply limited weathering) because higher heating rates cause more volcanism, and thus a higher concentration of planetary carbon in the crust, and higher temperatures, which lower the decarbonation depth (see \S \ref{sec:supply_lim}).  

Thus a large swath of the most favorable region of parameter space for long-lived CO$_2$ degassing would lead to inhospitably hot climates due to supply limited weathering, and is therefore uninhabitable (Figures \ref{fig:fd_time1}A \& B). Nevertheless, there is still a sizeable region of parameter space, at $Q_0 > 100$ TW and {$C_{tot} \sim 10^{20} - 10^{22}$ mol (corresponding to $\approx 7 \times 10^{-5} - 7 \times 10^{-3}$ mass \% CO$_2$)} where habitable climates can potentially be maintained for up to 2-4 Gyrs. Earth's total CO$_2$ budget is not well constrained, but estimates range from $\sim 10^{22} - 10^{23}$ mol \citep{Marty1998,Sleep2001b,Dasgupta2013}{, while the radiogenic heating budget is estimated at $\approx 70-100$ TW of Hadean heat production \citep[e.g.][]{korenaga2006}.} A planet with an Earth-like CO$_2$ and radiogenic heating budget would therefore be able to sustain a temperate climate for up to $1-2$ Gyrs. For longer maintenance of habitable conditions planets {with Earth's size and composition} would need larger radiogenic heating budgets than Earth and total CO$_2$ budgets similar to Earth's; larger CO$_2$ budgets would cause supply limited weathering.   

\begin{figure}
\includegraphics[width=0.75\textwidth]{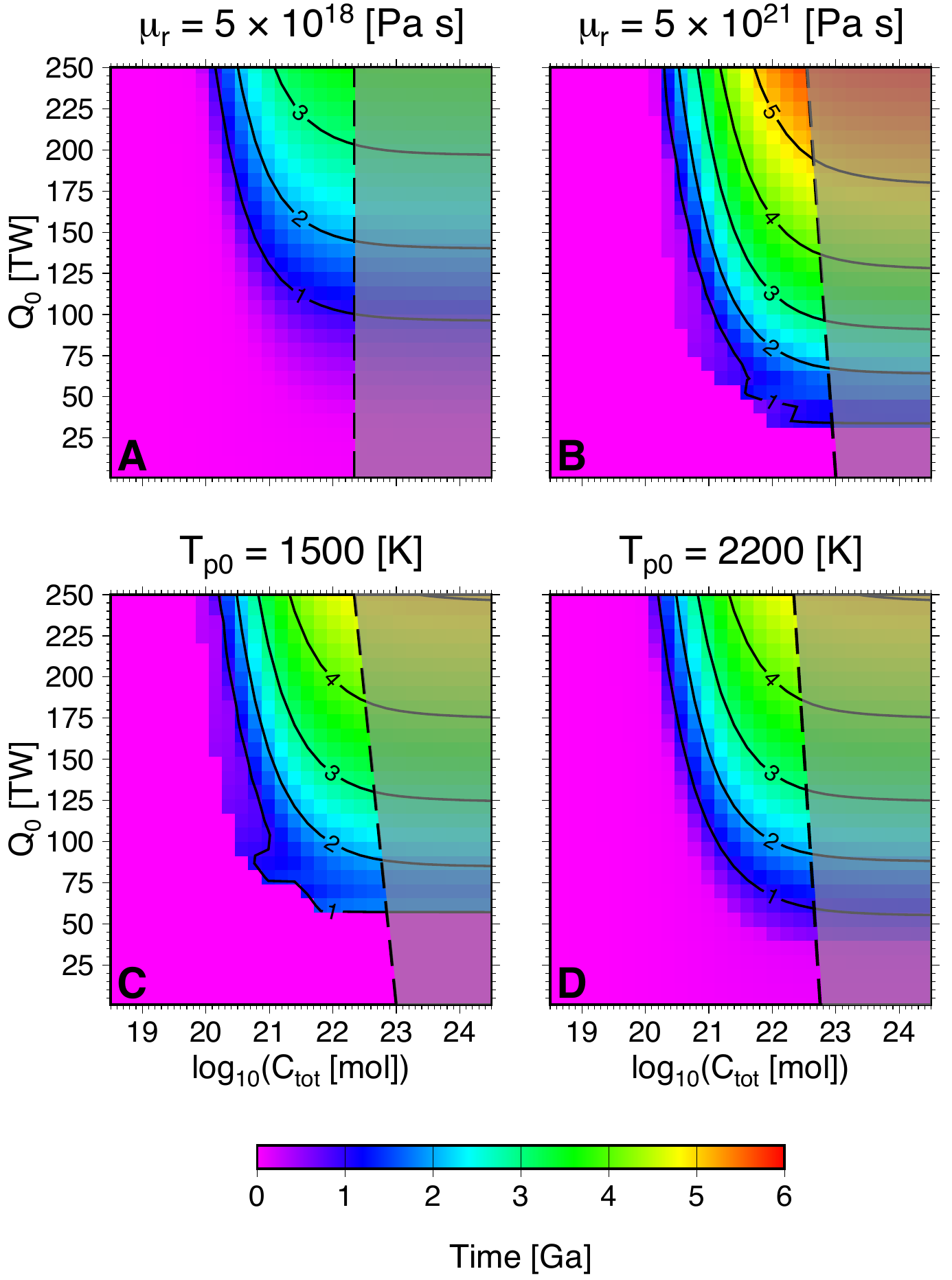}
\caption{\label {fig:fd_time2}  Time when the degassing flux falls below Earth's present day degassing flux for different model parameter choices: $\mu_r = 5 \times 10^{18}$ Pa s (A), $\mu_r = 5 \times 10^{21}$ Pa s (B), $T_{p_0} = 1500$ K (C), $T_{p_0} = 2200$ K (D). Dashed lines denote the transition to supply limited weathering as calculated from the thermal evolution models. {The supply limited weathering regime is shaded.} }
\end{figure} 

\subsection{Sensitivity of results to parameters and model formulation}
\label{sec:sensitivity}

Two key parameters that are not well constrained are $\mu_r$, the reference viscosity, and $T_{p_0}$, the initial mantle potential temperature. Varying these parameters over a reasonable uncertainty range leads to some noticeable differences, but does not significantly change the overall results (Figure \ref{fig:fd_time2}). A lower $\mu_r$ ($\mu_r$ is changed by changing the pre-exponential constant, $\mu_n$) shortens the lifetime of degassing, because the mantle convects more vigorously and cools more rapidly (Figure \ref{fig:fd_time2}A); planets with initial radiogenic heating budgets less than 100 TW see significant degassing and volcanism end in less than 1 Gyrs, while even at $Q_0 = 250$ TW degassing wanes in 3-4 Gyrs. Moreover, a lower $\mu_r$ also {shifts the onset of supply limited weathering} to lower $C_{tot}$, because more vigorous convection leads {to} a larger concentration of carbon in the crust, and {a} thinner lithosphere, and hence shallower decarbonation depth. Thus planets with low viscosity mantles are less likely to sustain habitable climates. A larger $\mu_r$ slows mantle cooling and allows degassing to last longer (Figure \ref{fig:fd_time2}B), but does not significantly expand the range of $C_{tot}-Q_0$ parameter space where long-lived degassing is possible as compared to the baseline model (Figure \ref{fig:fd_time1}). Lowering the initial mantle temperature suppresses volcanism at low radiogenic heating rates below $\approx 60$ TW, and therefore shrinks the region where habitable conditions are possible (Figure \ref{fig:fd_time2}C). However, for $Q_0 > 100$ TW there is no significant change to the degassing lifetime. Finally a larger initial mantle temperature does not change the calculated degassing longevities, but does {cause the onset of supply limited weathering to occur at} lower $C_{tot}$, as a hotter mantle leads to more outgassing and therefore concentration of carbon in the crust, and a lower decarbonation depth (Figure \ref{fig:fd_time2}D). 

\begin{figure}
\includegraphics[width=0.85\textwidth]{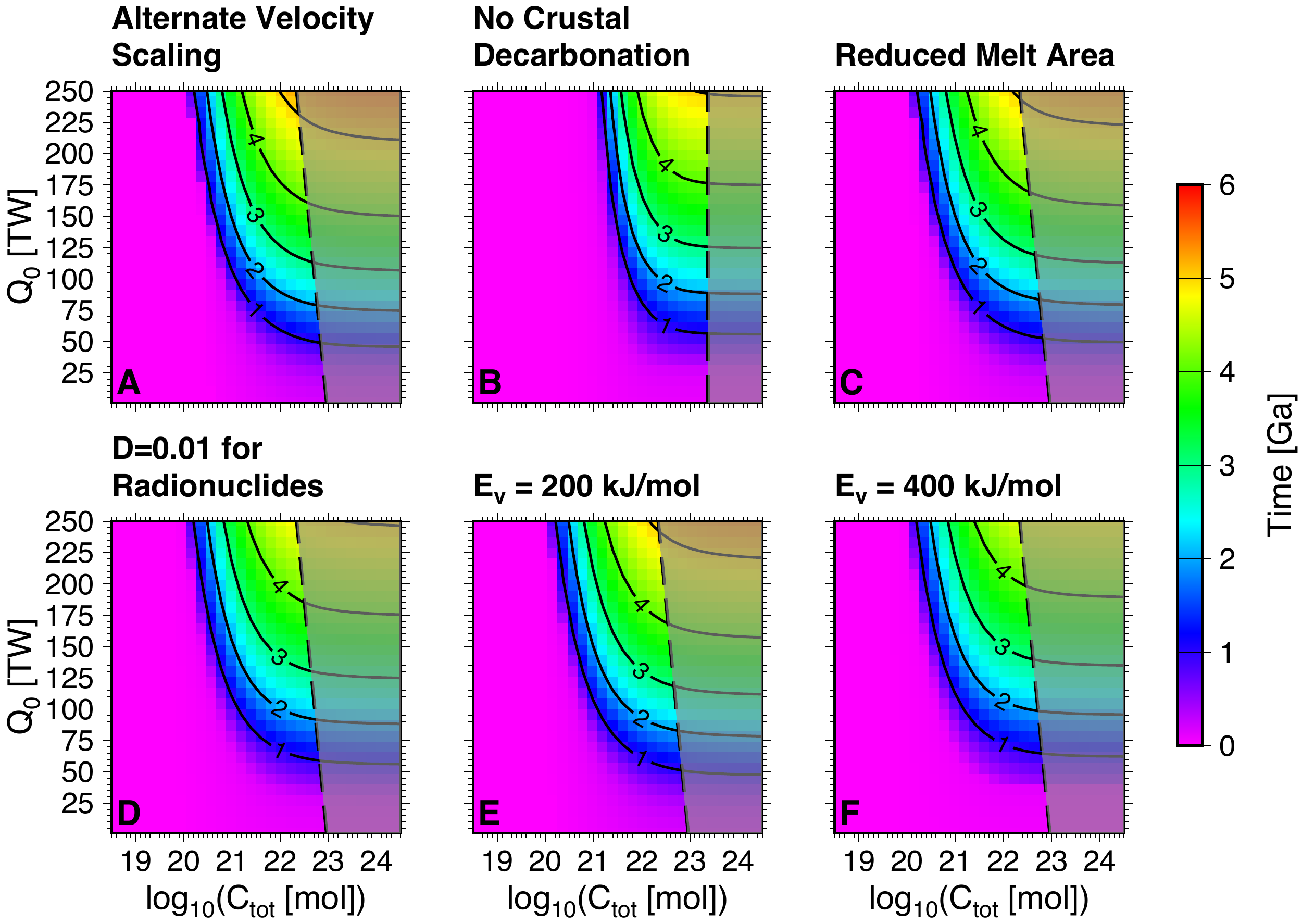}
\caption{\label {fig:fd_time3}  {Time when the degassing flux falls below Earth's present day degassing flux for: (A) using the velocity scaling law given by \eqref{vel2} instead of \eqref{vel1}; (B) assuming that $\delta_{carb}$ is always larger than $\delta$, such that no metamorphic decarbonation occurs within the crust; (C) assuming that upwelling mantle is confined to a cylinder with a radius half the size of the radius of convection cells; (D) using $D=0.01$ for heat producing elements; (E) using an activation energy for viscosity of $E_v = 200$ kJ mol$^{-1}$; and (F) using $E_v = 400$ kJ mol$^{-1}$. Shading denotes the supply limited weathering regime.}}
\end{figure}

Another key parameter is the eruption efficiency, $\epsilon$, or the fraction of melt produced that erupts at the surface. The eruption efficiency dictates the rate at which weatherable rock is supplied to the surface, and is therefore crucial for determining whether weathering becomes supply limited or not. The boundary { for the onset of supply limited weathering} scales linearly with $\epsilon$. Thus setting $\epsilon = 1$ shifts the transition to supply limited weathering to values of $C_{tot}$ one order of magnitude higher than those shown in Figures \ref{fig:fd_time1}-\ref{fig:fd_time3}, where $\epsilon = 0.1$, and gives the most optimistic possible estimate for when supply limited weathering would prevail on a stagnant lid planet. Choosing lower $\epsilon$ would shift the weathering regime boundary to lower $C_{tot}$. {Moreover, as discussed in \S \ref{sec:supply_lim}, the supply limit to weathering could be lower than we estimate if erupted basalt can not be completely carbonated. Incomplete carbonation would also shift the supply limited weathering regime to lower values of $C_{tot}$, in the same manner as lowering $\epsilon$. If only 1/100 or less of erupted basalt can be carbonated, then supply limited weathering will prevail over nearly the entire range of $C_{tot}$ and $Q_0$ where sufficient degassing rates take place; very inefficient crustal carbonation could thus prevent habitable climates on stagnant lid planets entirely.}

{Figure \ref{fig:fd_time3} shows the results of changing additional key aspects of the model.} Changing the convective velocity scaling law from equation \eqref{vel1} to equation \eqref{vel2} (i.e. from a scaling exponent of $2/3$ to $1/2$){, causes degassing lifetimes to be $\approx 500$ Myrs longer overall than our baseline model results where \eqref{vel1} is used (Figure \ref{fig:fd_time3}A).} The reason degassing and volcanism lasts longer {with a} $Ra_i^{1/2}$ scaling law {is that convective velocities are lower, resulting in} lower melt production rates, and thus less depletion of heat producing elements in the mantle. However, the difference between the two velocity scaling laws on degassing lifetime and the transition to supply limited weathering is minor compared to other sources of uncertainty in the modeling. {We also test the effect of reducing the melt production rate, $f_m$, by a factor of 2 as compared to our standard formulation given in \eqref{fm}, as explained in \S \ref{sec:melting}. The results are nearly identical to our baseline model, and actually show longer lived degassing as a result of greater retention of heat producing elements in the mantle (Figure \ref{fig:fd_time3}C).}

{We also test the effect of the distribution coefficient, $D$, which controls the degree to which heat producing elements are preferentially partitioned into the crust during melting, and the activation energy for viscosity, $E_v$.} The upper limit {for the distribution coefficient} is $D \approx 0.01$ for a mantle dominated by pyroxene \citep{Hart1993,Hauri1994}. Using $D=0.01$ produces no discernible differences in the lifetime of CO$_2$ degassing {(Figure \ref{fig:fd_time3}D).} Distribution coefficients of $D > \sim 0.1-1$ are needed to significantly impede enrichment of heat producing elements in the crust, and produce longer degassing lifetimes than calculated in our baseline models. {Varying the activation energy for viscosity by 100 kJ mol$^{-1}$ above and below our baseline value does not change the region of $C_{tot}-Q_0$ space where potentially habitable climates are possible, but does change the lifetime of sufficient CO$_2$ degassing; with $E_v = 200$ kJ mol$^{-1}$ degassing lasts longer than in the baseline model, while with $E_v = 400$ kJ mol$^{-1}$ it ends sooner. The stronger the dependence of viscosity on temperature, the more lid thickness grows with decreasing mantle temperature. Thus a higher activation energy causes the lid to grow more rapidly as the mantle cools and shut off volcanism, and outgassing, sooner. Note that when we change activation energy, the reference viscosity, $\mu_r$, is held constant; this is accomplished by changing $\mu_n$ when $E_v$ is changed. For $E_v = 200$ kJ mol$^{-1}$ $\mu_n = 7.3 \times 10^{13}$ Pa s, and for $E_v = 400$ kJ mol$^{-1}$ $\mu_n = 2.7 \times 10^7$ Pa s. }

{Finally, we test how our results are affected if metamorphic decarbonation does not occur. Considering a case where crust does not decarbonate as it is buried is useful for two main reasons: 1) Exoplanets can have a wide range of chemical compositions, and crustal decarbonation may not occur at the temperature and pressure conditions prevalent in the crust on all planets. We note, however, that our model assumes an olivine dominated mantle and basaltic crust, and is thus not directly applicable to planets with different mineralogies. 2) Rapid burial of the crust could suppress temperatures in the lid by advecting cold rocks from the surface downward faster than thermal diffusion can heat them to the steady-state geotherm. For melt production rates typically seen in our models, advection would only have a minor influence on the temperature profile in the crust as $Pe < 1$ (\S \ref{sec:thermal_evol}). However, if volcanism is confined to only a handful of places on the surface local crustal burial rates would be higher, and crust could potentially founder into the mantle before it heats enough to undergo decarbonation. Figure \ref{fig:fd_time3}B shows} that the range of $C_{tot}$ that can support long-lived degassing shifts to higher values when crustal decarbonation is no longer active. However, the supply limit to weathering also shifts to larger $C_{tot}$, meaning that the size of the region of parameter space where long-lived habitable climates are possible is approximately the same, regardless of whether metamorphic decarbonation occurs or not. {Without metamorphic decarbonation the onset of supply limited weathering shifts to higher values because} the planet's CO$_2$ supply largely remains in the mantle, and the large volume of the mantle means this carbon is diluted. Thus the same $C_{tot}$ results in an overall lower degassing rate when only mantle volcanism is active, than when both mantle volcanism and crustal decarbonation can take place. 

\subsection{Erupted melt fluxes and habitability}
\label{sec:melt_flux}

\begin{figure}
\includegraphics[width=0.5\textwidth]{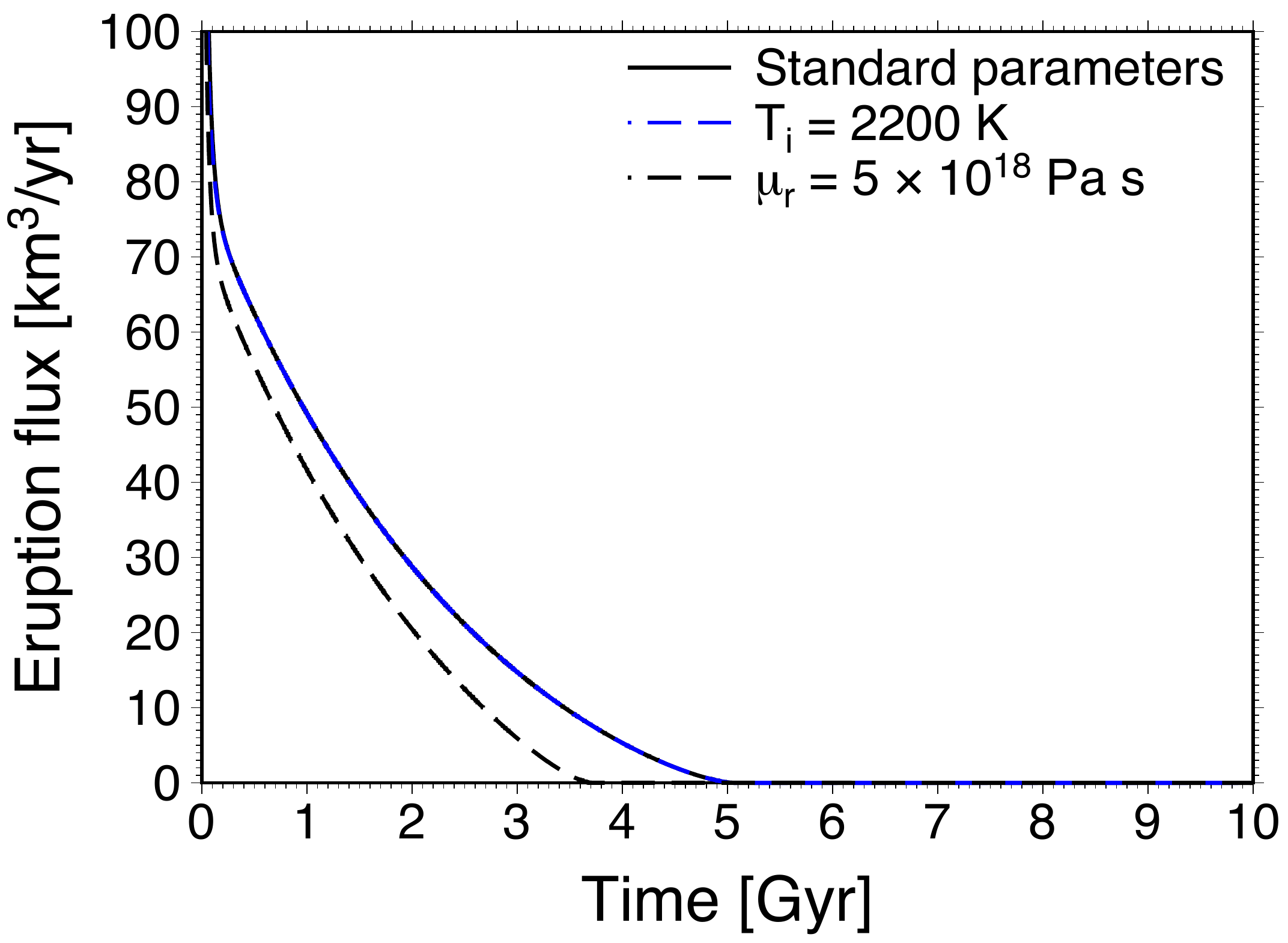}
\caption{\label {fig:eflux}  Volume flux of erupted magma for thermal evolution models with $Q_0 = 250$ TW, the highest initial heating rate considered in this study. The solid black line gives the eruption flux for a model with the standard parameter choices (Table \ref{tab_param}), the dashed blue line assumes an initial mantle temperature of $T_i = 2200$ K (note the dashed blue line sits almost directly on top of the solid black line), and for the dashed black line $\mu_r = 5 \times 10^{18}$ Pa s. Results shown are those with the largest eruption fluxes for the suite of models performed in this study.}
\end{figure} 

Although active volcanism is beneficial for habitability to the extent that it promotes a temperate climate, extreme volcanic resurfacing rates could prevent life from developing by rapidly sterilizing the surface. {The resurfacing rate at which life would be detrimentally impacted by lava flows is not well known. Despite this uncertainty, we can at least compare our eruption rates to those seen on Earth. Eruption rates are calculated from our thermal evolution models as $\epsilon f_m$ with $\epsilon = 0.1$ (Figure \ref{fig:eflux}). Eruption rates are typically $< 100$ km$^{3}$ yr$^{-1}$ for most of the period while volcanism is active, which results in $\approx 0.2$ mm yr$^{-1}$ of resurfacing when erupted lava is assumed to be distributed globally. For comparison, resurfacing rates at Kilauea Volcano in Hawaii are $\approx 5$ mm yr$^{-1}$ \citep[e.g.][]{Quane2000}, and life is present across the volcano. Thus eruptions are unlikely to be a major impediment to life on stagnant lid planets for most of the period when volcanism is active. Initial volcanic rates are often very high (on the order of $10^4$ km$^3$ yr$^{-1}$), and potentially uninhabitable, in our models, but such high volcanic rates are short lived (lasting on the order of 100 Myrs).  }
      

\section{Discussion}
\label{sec:discussion}

\subsection{Implications for exoplanet observations and target selection}

The model results indicate that Earth-sized stagnant lid planets with {extrapolated initial} radiogenic heating rates {greater than Earth's,} and with total CO$_2$ budgets of {$\sim 10^{20} - 10^{22}$ mol ($\approx 7 \times 10^{-5} - 7 \times 10^{-3}$ mass \% CO$_2$),} are potentially habitable for timescales of over 2 billion years. {The time when habitability becomes unlikely due to limited CO$_2$ outgassing scales with initial radiogenic heat budget, with higher budgets leading to longer-lived outgassing. Thus planet heat budget, CO$_2$ budget, and age are critical factors for habitability of stagnant lid planets.} Whether planets falling within these limits are common or not is not well constrained, in particular regarding CO$_2$ inventories. Accreted CO$_2$ inventories are likely strongly dependent on the size and composition of the proto-planetary disk, and the ability of volatile rich planetesimals to be scattered inward towards the terrestrial planet formation zone, likely by interaction with giant planets like Jupiter \citep{Raymond2004}. When volatile rich planetesimals are scattered inward significant radial mixing occurs, with water and other volatiles delivered to terrestrial planets in a highly stochastic manner. In models of solar system formation, \cite{Raymond2004} found planets formed with water concentrations ranging from 2 orders of magnitude lower than Earth's to two orders of magnitudes larger, with the majority of planets falling within {a range of $\sim 1-100$ ocean masses of water (i.e. $\sim 0.1-1$ to $\sim 10-100$ times Earth's total water abundance).} CO$_2$ delivery is likely similar, as it would also be caused by inward scattering of volatile rich planetesimals, meaning stochastic processes probably lead to a similar range of final planetary CO$_2$ inventories. Ultimately it is difficult to predict or constrain from stellar observations which exoplanet systems are likely to have planets with bulk CO$_2$ budgets within the optimum range for habitability indicated by our models.

Radiogenic heating budget and age, on the other hand, can be constrained by stellar observations, {and both stars with sufficiently high radionuclide abundances, and those young enough to potentially host habitable stagnant lid planets, are apparently common.} 
\cite{Unterborn2015} showed that the thorium abundance of sun-like stars varies from 59 \% to 251 \% the solar value, {for stars with abundances of major rock forming elements that would produce planet mineralogies similar to Earth. These thorium variations} would lead to radiogenic heating budgets ranging from $\approx 0.5 - 2.5$ times Earth's. {Moreover,} most stars measured in \cite{Unterborn2015} had higher thorium abundances than the sun. {Planet age can be inferred from star age, measurable by astroseismology \citep[e.g.][]{SilvaAguirre2015,Burgasser2017}, and results from Kepler stars show many planet hosting stars with ages of 2-4 Gyrs \citep{SilvaAguirre2015}. 
Thus the likelihood of planets meeting the CO$_2$, radiogenic heating budget, and age constraints for maintaining} habitable climates in the stagnant lid regime appears to be relatively high. As a result, plate tectonics may not be required for habitability, although as we discuss in \S \ref{sec:pt}, it still has some major benefits compared to stagnant lid convection. Furthermore, in order to avoid a fate similar to Venus, stagnant lid planets would still need to lie within the habitable zone, and in order to avoid becoming barren, frozen worlds like Mars, would need to be large enough for the planet's gravity to prevent rapid escape of atmospheric greenhouse gases. 

{Our results can also naturally be used to guide target selection for future missions. In general, the model results indicate that young planets and/or those which orbit stars with high U and Th abundances are good targets for searching for biosignatures, as these planets are potentially habitable even in a stagnant lid regime. The modeling presented here could even be used in future applications to directly constrain the prospects for habitability of individual planets as they are detected.}
Whether high heat production rates also enhance a planet's prospects for plate tectonics is not clear, however, as how mantle temperature and heat production rate influence a planet's propensity for plate tectonics is debated \citep{ONeill2007b,Korenaga2010a,Foley2014_initiation,Foley2016_review}. Another potential way to use the results for target selection is to estimate the amount of tidal heating that would be expected on exoplanets based on their orbits. Planets experiencing significant tidal heating will be able to maintain long-lived CO$_2$ degassing, even if the abundance of heat producing elements in the mantle is low. Tidal heating is {potentially} important for rocky planets in the habitable zone of M-dwarfs \citep{Jackson2008}, or planets or moons undergoing orbital interactions with other planets, similar to Io \citep{Moore2003}. {In particular, stagnant lid planets orbiting M-dwarfs at the outer edge of the habitable zone can see significant tidal heating last for $\sim 10$ Gyrs, before their orbits circularize \citep{Driscoll2015b}; circularization is much faster for planets near the inner edge of the habitable zone.} 

\subsection{Comparison to plate-tectonic planets}    
\label{sec:pt}

On a plate-tectonic planet CO$_2$ degassing is easier to maintain. Ridges bring hot mantle all the way to the surface, meaning that volcanism at divergent plate margins will continue until the mantle potential temperature drops below the zero pressure solidus of $\approx 1400$ K \citep{Herzberg2000}. Plate tectonics also allows for continual metamorphic degassing through subduction. Slab carbon loss is most likely driven by dissolution in fluids released by metamorphic dehydration reactions \citep{Ague2014}, as temperature and pressure conditions along modern slab top geotherms do not cause significant CO$_2$ release \citep{Kerrick2001,Kerrick2001b}. Thus arc degassing will continue as long as significant slab dewatering and fluid-flux melting of the mantle wedge occurs. Hydrous wedge melting takes place at low temperatures of $\approx 900-1100$ K \citep{Hacker2003}. For mantles cooler than this, all slab carbon will be recycled into the mantle, but without slab or ridge degassing CO$_2$ release to the atmosphere will be negligible and the surface will likely be frozen. How long it takes the mantle of a plate-tectonic planet to cool below these temperature thresholds depends on the details of an individual planet's evolution. However, \cite{Kite2009} provides a reasonable estimate based on simple models that planetary mantles will stay warm enough for ridge degassing for over 10 Gyrs. Slab degassing can be maintained at lower temperatures and would thus last even longer.

Plate tectonics also allows {a stabilizing weathering feedback}, and thus temperate climates, to be maintained at larger {total} CO$_2$ budgets than stagnant lid planets when the area of exposed land is $> \sim 1$ \% of the plate tectonic planet's surface \citep{Foley2015_cc}. A large land area, combined with the high erosion rates typical of mountainous and tectonically active regions, produces a high supply limit to weathering, which requires a very large total degassing flux in order to be reached. Thus a plate-tectonic planet can potentially maintain a temperate climate at CO$_2$ inventories up to $\sim 10^{24}-10^{25}$ mol, much larger than the $\sim 10^{23}$ mol upper limit for stagnant lid planets when crustal decarbonation is active. With little or no land, the CO$_2$ budget where weathering becomes supply limited on a plate-tectonic planet is approximately equal to the estimates shown here for stagnant lid planets with crustal decarbonation. 

\subsection{Uncertainties in model formulation}

Although some differences can be seen, the overall results, in particular the size of the region of $C_{tot}-Q_0$ parameter space where habitable climates are possible, are {consistent} over a wide range of parameters, and to changes in the formulation for convective velocity, {metamorphic decarbonation of the crust, melt transport, and volcanic outgassing.} Nevertheless, there are effects not included in this study that could influence the results. The most obvious is size, as many rocky exoplanets more massive than Earth have been discovered \citep[e.g.][]{Batalha2014}. {Larger planets have a smaller surface area to volume ratio, and thus retain their heat for longer, all else being equal. Therefore increasing planet size is expected to prolong mantle melting and volcanism.} However, melt will be formed at higher pressures due to the larger pressure gradient on a massive planet, and could end up being too dense to erupt. \cite{Noack2017} argue that planets larger than $\approx 2-3$ Earth masses will experience no outgassing because dense melt will prevent eruption. Future studies of melting and melt density at super-Earth conditions are needed to confirm this result, but it is possible that there is a critical planet size above which a stagnant lid planet would not be habitable. 

{Considering only Earth-sized planets, there are additional important uncertainties to highlight. Our assumptions} about melt heat loss maximize this effect, and thus hasten the end of volcanism. A more realistic model, particularly regarding the degree to which melt solidifying at shallow depth contributes to net mantle cooling, may allow degassing to be sustained longer, and over a wider range of parameter space, than presented here. We also ignore melt liberation of CO$_2$ locked in the crust, which would enhance CO$_2$ outgassing during mantle volcanism. Melt liberation of crustal carbon is a potentially significant CO$_2$ source on Earth \citep{Lee2013}, and including it would allow habitable conditions at lower {total} CO$_2$ budgets. Thus, the true prospects for habitability on stagnant lid planets {could be} better than our results indicate. Core cooling would also lead to the formation of plumes, which could sustain volcanism even after pressure release melting due to passive mantle upwelling has ceased. However, plumes alone may not be a sufficient source of volcanism to sustain a temperate climate. {Plumes on the modern Earth result in $\sim 1-3$ km$^3$ yr$^{-1}$ of melt flux \citep{Crisp1984}, which is similar to the melt fluxes seen in our models when volcanism is close to ceasing and CO$_2$ outgassing levels are too low to prevent glaciation.} The effect of mantle volatile content on viscosity and the solidus is also not included. The viscosity of wet olivine is 1-2 orders of magnitude lower than dry olivine \citep{Hirth2003}, meaning that the mantle could become more viscous over time as water is outgassed to the surface. {At the same time, the solidus will shift from lower to higher temperatures as mantle water content declines \citep[e.g.][]{Katz2003}.} As a result, the mantle cooling rate would decline as water is expelled from the mantle, {acting to keep the mantle hotter, but the solidus temperature will also increase. New models would be needed to determine which of these competing effects of mantle hydration state on CO$_2$ outgassing are dominant.} 

{There are also factors that could make habitability less likely than our results indicate. First, as discussed in \S \ref{sec:supply_lim} \& \S \ref{sec:sensitivity}, we assume a supply limit to weathering based on complete carbonation of basalt. In reality, the amount of basalt that can be carbonated will be controlled by the ability of pore fluids to reach fresh mineral surfaces. The competition between reaction-driven cracking, acting to open up new pore space, versus permeability loss due to mineral precipitation, will determine the amount of basalt that can be carbonated \citep[e.g.][]{Rudge2010}. 
As a result, only a fraction of surface basalt may actually be available for weathering. If only $\sim 1/100$ of surface basalt can be carbonated, then nearly all of the parameter space where long-lived degassing is possible will result in supply limited weathering, and habitability of stagnant lid planets will be unlikely. Metamorphic degassing may also be less significant than we model, or even entirely absent, if eruption is concentrated in a few localized areas such that crustal burial at these sites is rapid. However, we show that even without metamorphic decarbonation there is still a sizable region of $C_{tot}-Q_0$ parameter space where habitable climates are possible (see \S \ref{sec:sensitivity}).  


We also assume complete outgassing of all melt produced to the atmosphere. However, a significant fraction of melt generated could stall in the crust or stagnant lid and not outgas to the atmosphere, potentially making it more difficult than our models predict to sustain degassing rates high enough to avoid a snowball climate. Simple test cases designed to capture the effect of melt stalling and incomplete degassing are presented in Appendix \ref{sec:melt_stall}, and show no significant difference from our models where all the melt degasses. Finally, we assume that volcanism and outgassing are continuous in time, as melt is constantly being created by mantle upwelling into the melting region. Continuous volcanism is important because pauses in CO$_2$ outgassing lasting longer than $\sim 10^5$ yrs could trigger snowball states \citep{Tajika2008}. While melt is continuously being generated in the mantle in our models, it is possible that melt transport through the lid could induce episodicity in eruptions that our simple thermal evolution model can not capture. Ultimately future work involving convection models and models of melt transport through a stagnant lid will be needed to test whether significant pauses in stagnant lid planet volcanism are likely, and the effect they would have on habitability.}

\section{Conclusions}
\label{sec:conclusions}

Models of the thermal, magmatic, and degassing history of rocky planets {with Earth-like size and composition} demonstrate that a carbon cycle capable of regulating atmospheric CO$_2$ content, and stabilizing climate to temperate surface temperatures, can potentially operate on geologic timescales on planets in the stagnant lid regime. Plate tectonics may not be required for habitability, at least in regards to sustaining a stable, temperate climate on a planet. Specifically, stagnant lid planets with radiogenic heating budgets similar to or larger than Earth's, and with total CO$_2$ budgets of {$\sim 10^{20}-10^{22}$ mol (ranging from $10^{-2} - 1$ times typical estimates for Earth's CO$_2$ budget)} can potentially maintain temperate climates for 1-5 Gyrs. After this time volcanism and CO$_2$ outgassing largely cease, and climate would likely cool below the water freezing point and induce global surface glaciation {(though plumes, which are neglected in our modeling, could prolong volcanism and temperate climates)}. At CO$_2$ budgets lower than {$\sim 10^{20}$ mol} a planet's climate is estimated to be in a snowball state for its entire history, while above {$\sim 10^{22}$} mol weathering would become supply limited. With supply limited weathering CO$_2$ outgassing overwhelms CO$_2$ drawdown, such that an inhospitably hot, CO$_2$ rich atmosphere forms. Thus the amount of carbon accreted to a planet during formation is critical for whether it can sustain habitable surface conditions in a stagnant lid regime. Our results also indicate that high internal heating rates favor long-term habitability by prolonging volcanism and CO$_2$ degassing{, and that young planets are more likely to experience sufficiently high rates of CO$_2$ outgassing than older planets.} Thus, planets orbiting high thorium or uranium abundance stars {or young stars} are good targets for searching for biosignatures, because they are more likely to be habitable, even if the mode of surface tectonics is stagnant lid rather than plate tectonics.   

\section{Author disclosure statement and acknowledgements}

No competing financial interests exist. {We thank Ed Kite, Lena Noack, and two anonymous reviewers whose comments helped us significantly improve the manuscript.} 

\appendix
{
\section{Appendix: Details of modeling phase equilibria}
\label{sec:appendix}
}
The following solution models were used in the computation of phase relations (Figure \ref{fig:phase}); all other phases considered were treated as ideal:

Chlorite: Chl(HP) -  \citep{Holland1998}

Biotite: Bio(HP) - \citep{Powell1999}

Garnet: Gt(HP) -  \citep{Holland1998a}

Amphibole: GlTrTsMg, GlTrTsPg - \citep{Wei2003,White2001,White2003} 

Clinopyroxene: Cpx(HP) - \citep{Holland1996}

Epidote: Ep(HP) - \citep{Holland1998a}

CO$_2$-H$_2$O fluid: \citep{Holland1991}

Magnesite: M(HP) - \citep{Holland1998a}

Orthopyroxene: Opx(HP) -  \citep{Holland1996}

Dolomite: Do(HP) - \citep{Holland1998a}

Phengite: Pheng(HP) - \citep{Holland1998a}

Spinel: Sp(HP) - \citep{Holland1998a} 

{
\section{Appendix: Effect of melt stalling and incomplete degassing}
\label{sec:melt_stall}

In the thermal evolution and outgassing models presented in the main text, all of the melt produced is assumed to percolate to the surface or near surface. As such, all of the CO$_2$ in the melt is assumed to degas to the atmosphere. Moreover, all CO$_2$ liberated from the crust by metamorphic reactions is also assumed to degas to the atmosphere. In this appendix we present test cases where degassing of CO$_2$ from the melt and from metamorphic decarbonation is incomplete, and where melt is allowed to stall at the base of the stagnant lid and not contribute to surface volcanism and outgassing. These tests assess the sensitivity of our results to the assumption of complete melt percolation to the surface and complete outgassing of CO$_2$ from the melt and metamorphic reactions. 

We first consider a case where all of the melt is still assumed to rapidly rises to the surface, but that only a fraction of the CO$_2$ in this melt degasses; the rest is trapped within the crust and becomes part of the crustal CO$_2$ reservoir. As a result, the volcanic degassing flux, $F_d$, is reduced by a factor $\zeta_1$: 
\begin{equation}
F_d = \frac{\zeta_1 f_m R_{man} [1 - (1-\phi)^{1/D_{CO_2}} ]}{\phi (V_{man}+V_{lid})}.
\end{equation}  
We also allow for a fraction, $\zeta_2$, of the CO$_2$ released by metamorphic reactions to remain trapped in the crust, and eventually recycle into the mantle. As a result, the metamorphic degassing flux is 
\begin{equation}
F_{meta} = (1-\zeta_2) \frac{R_{crust}f_m}{2 V_{carb}}(\tanh{((\delta_c-\delta_{carb})20)}+1) 
\end{equation}
and the evolution of the mantle and crustal carbon reservoirs follow 
\begin{equation}
\begin{split}
\frac{dR_{man}}{dt} & = -\frac{f_m R_{man} [1 - (1-\phi)^{1/D_{CO_2}} ]}{\phi (V_{man}+V_{lid})} + \\ & \zeta_2 \frac{R_{crust}}{V_{crust}}\left(f_m-4\pi(R_p-\delta)^2\textrm{min}\left(0,\frac{d\delta}{dt}\right) \right)(\tanh{((\delta_c-\delta)20)}+1)
\end{split}
\end{equation} 
\begin{equation}
\begin{split}
\frac{dR_{crust}}{dt} & = \frac{f_m R_{man} [1 - (1-\phi)^{1/D_{CO_2}} ]}{\phi (V_{man}+V_{lid})} - \\ & \zeta_2 \frac{R_{crust}}{V_{crust}}\left(f_m-4\pi(R_p-\delta)^2\textrm{min}\left(0,\frac{d\delta}{dt}\right) \right)(\tanh{((\delta_c-\delta)20)}+1)
\end{split}
\end{equation} 
when $\delta_{carb} < \delta$, and \eqref{rman2} and \eqref{rcrust2} otherwise. Note that transfer of CO$_2$ from the mantle to the crust is the same as in the main text, as all CO$_2$ that is stripped from the mantle by melting is assumed to be deposited in the crust regardless of whether it outgases into the atmosphere first or is trapped in the crust during solidification. The rest of the thermal evolution and carbon cycle model is the same as presented in the main text. 

The results of test cases with $\zeta_1 = 0.1$ and $\zeta_2 = 0$, and with $\zeta_1 = 0.01$ and $\zeta_2 = 0$ are identical to the results of the baseline model in the main text (Figure \ref{fig:fd_time4}A \& B). As it is the metamorphic decarbonation that dominates outgassing for most of the lifetime of volcanism in our models, incomplete degassing of the melt does not significantly change the results. The results are changed when $\zeta_2 > 0$ is used; specifically as $\zeta_2$ approaches 1, model results converge to the case where crustal decarbonation is ignored (i.e. Figure \ref{fig:fd_time3}B in the main text), as $\zeta_2=1$ implies no metamorphic CO$_2$ outgassing, and hence all of the crustal carbon being recycled into the mantle by foundering. As a result, models with $\zeta_1=0.1$ and $\zeta_2 = 0.9$ (i.e. where only 10 \% of the carbon that is metamorphically lost from the crust degasses to the atmosphere), produce very similar degassing lifetimes to the no crustal decarbonation case in the main text (Figure \ref{fig:fd_time4}C). However, because volcanic outgassing is also assumed to be incomplete, overall degassing rates are lower and both the transition to supply limited weathering and the range of total CO$_2$ budgets where long-lived degassing occurs shift to higher values. Thus the range of $C_{tot}$ where habitable climates are possible is approximately the same size as in the models presented in the main text, but shifted to higher values as a result of incomplete outgassing of both melt and CO$_2$ released by metamorphism.   

Another important scenario to test is one where some fraction of melt produced in the mantle cannot rise through the stagnant lid and reach the surface or near surface. We therefore perform models where only a fraction, $\xi$, of the melt produced reaches the surface and contributes to mantle cooling, crustal growth, and CO$_2$ outgassing. The rest is assumed to be trapped at the base of the lid, and therefore remains in the mantle. The model is thus modified as follows. The mantle energy balance becomes 
\begin{equation}
V_{man} \rho c_p \frac{dT_p}{dt} = Q_{man} - A_{man} F_{man} - \xi f_m \rho_m \left (c_p \Delta T_m + L_m \right) .
\end{equation} 
Heat producing elements in the crust and mantle evolve as   
\begin{equation}
\begin{split}
\frac{d Q_{crust}}{dt} & = \frac{x_m \xi f_m}{\phi} [1 - (1-\phi)^{1/D} ] - \\ & x_c \left(\xi f_m-4\pi(R_p-\delta)^2\textrm{min}\left(0,\frac{d\delta}{dt}\right) \right)(\tanh{((\delta_c-\delta)20)}+1) - \frac{Q_{crust}}{\tau_{rad}}
\end{split}
\end{equation} 
\begin{equation}
\begin{split}
\frac{d Q_{man}}{dt} & = x_c \left(\xi f_m-4\pi(R_p-\delta)^2\textrm{min}\left(0,\frac{d\delta}{dt}\right) \right)(\tanh{((\delta_c-\delta)20)}+1) - \\ & \frac{x_m \xi f_m}{\phi} [1 - (1-\phi)^{1/D} ] - \frac{Q_{man}}{\tau_{rad}}
\end{split}
\end{equation} 
and the mantle and crustal carbon reservoirs follow 
\begin{equation}
\begin{split}
\frac{dR_{man}}{dt} & = -\frac{\xi f_m R_{man} [1 - (1-\phi)^{1/D_{CO_2}} ]}{\phi (V_{man}+V_{lid})} + \\ & \zeta_2 \frac{R_{crust}}{V_{crust}}\left(\xi f_m-4\pi(R_p-\delta)^2\textrm{min}\left(0,\frac{d\delta}{dt}\right) \right)(\tanh{((\delta_c-\delta)20)}+1)
\end{split}
\end{equation} 
\begin{equation}
\begin{split}
\frac{dR_{crust}}{dt} & = \frac{\xi f_m R_{man} [1 - (1-\phi)^{1/D_{CO_2}} ]}{\phi (V_{man}+V_{lid})} - \\ & \zeta_2 \frac{R_{crust}}{V_{crust}}\left(\xi f_m-4\pi(R_p-\delta)^2\textrm{min}\left(0,\frac{d\delta}{dt}\right) \right)(\tanh{((\delta_c-\delta)20)}+1)
\end{split}
\end{equation} 
when $\delta_{carb} < \delta$, and
\begin{equation}
\begin{split}
\frac{dR_{crust}}{dt} & =\frac{\xi f_m R_{man} [1 - (1-\phi)^{1/D_{CO_2}} ]}{\phi (V_{man}+V_{lid})}- \\ & \frac{R_{crust}}{V_{crust}}\left(\xi f_m-4\pi(R_p-\delta)^2\textrm{min}\left(0,\frac{d\delta}{dt}\right) \right)(\tanh{((\delta_c-\delta)20)}+1)
\end{split}
\end{equation} 
\begin{equation}
\begin{split}
\frac{dR_{man}}{dt} & = -\frac{\xi f_m R_{man} [1 - (1-\phi)^{1/D_{CO_2}} ]}{\phi (V_{man}+V_{lid})}+ \\ & \frac{R_{crust}}{V_{crust}}\left(\xi f_m-4\pi(R_p-\delta)^2\textrm{min}\left(0,\frac{d\delta}{dt}\right) \right)(\tanh{((\delta_c-\delta)20)}+1)
\end{split}
\end{equation}
when $\delta_{carb} > \delta$. 

\begin{figure}
\includegraphics[width=0.85\textwidth]{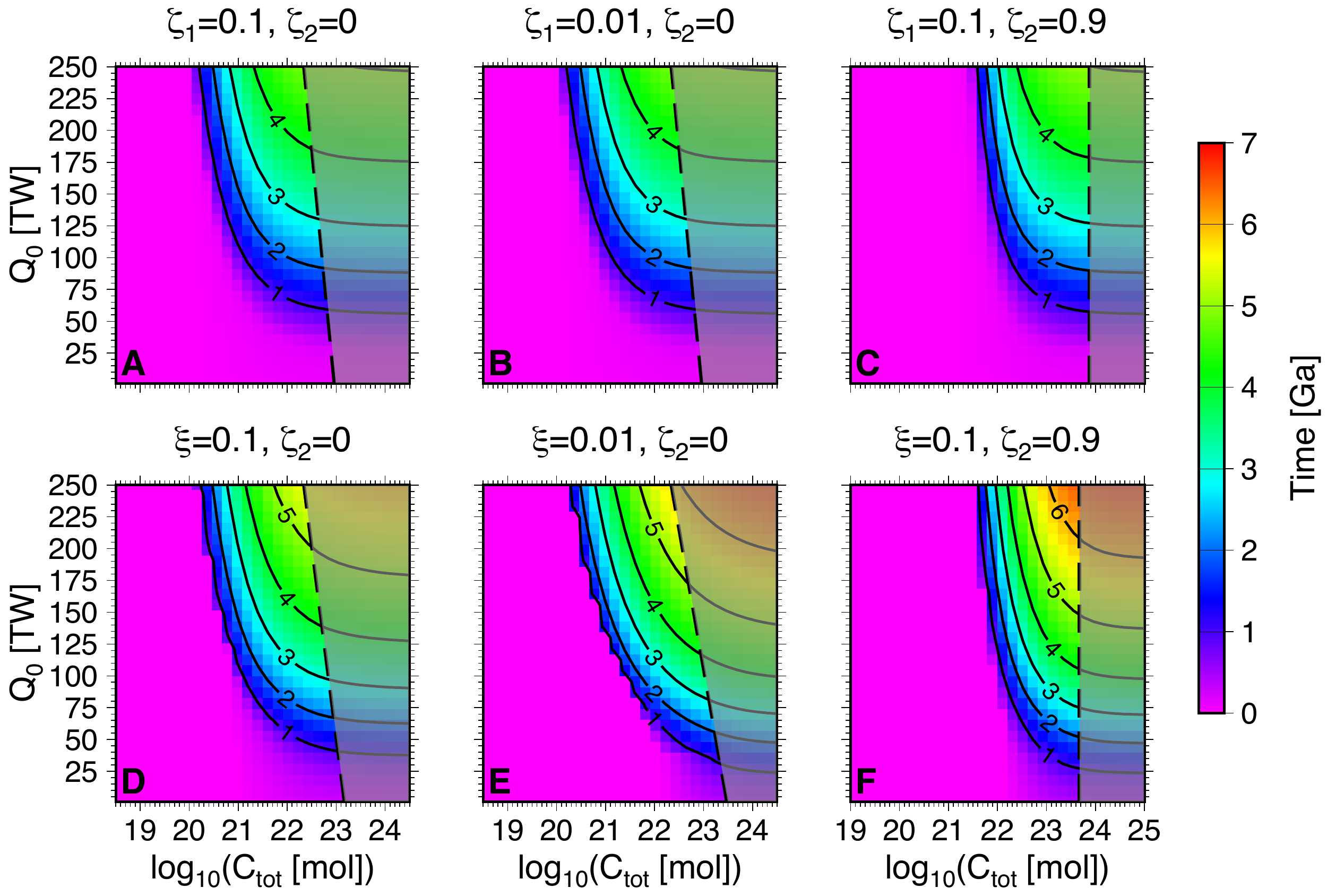}
\caption{\label {fig:fd_time4}  {Time when the degassing flux falls below Earth's present day degassing flux for models with incomplete degassing and melt transport through the lid: (A) $\zeta_1=0.1$ and $\zeta_2=0$; (B) $\zeta_1=0.01$ and $\zeta_2=0$; (C) $\zeta_1=0.01$ and $\zeta_2=0.9$; (D) $\xi=0.1$ and $\zeta_2=0$; (E) $\xi=0.01$ and $\zeta_2=0$; and (F) $\xi=0.1$ and $\zeta_2=0.9$. The supply limited weathering regime is shaded.}}
\end{figure}  

With melt stalling at the base of the lid, the model results are different than those presented in the main text in detail, but the range of $C_{tot}$ and $Q_0$ that allows for potentially habitable climates is nearly identical (Figure \ref{fig:fd_time4}D \& E). In fact, the main difference when melt is assumed to stall at the base of the lid is that degassing lifetimes are everywhere longer; the mantle cools more slowly with less efficient melt heat loss and fewer radionuclides partitioning into the crust. When metamorphic decarbonation is also assumed to be limited by setting $\zeta_2 = 0.9$, we obtain results similar to the case in the main text where decarbonation is excluded (Figure \ref{fig:fd_time4}F). As explained above, assuming that the majority of metamorphically liberated CO$_2$ is unable to outgas is effectively the same as assuming the crust does not undergo decarbonation. Again, degassing lifetimes are found to be longer with melt stalling because the mantle loses less heat from volcanism and fewer heat producing elements to the crust. Ultimately, our test cases demonstrate that the results in the main text are robust to the effects of incomplete outgassing of CO$_2$ and melt stalling in the lower lithosphere, at least as far as these affects can be incorporated into our simple thermal evolution models. Ultimately more sophisticated modeling involving two phase flow of melt and CO$_2$ vapor through the solid lithosphere is needed to more thoroughly assess how much melt produced in the mantle is actually able to degas at the surface, and the effect this has on long-term climate evolution. }

\clearpage



\end{document}